\documentclass[english,aps,pre,amsmath,amssymb]{revtex4}
\usepackage[utf8x]{inputenc}
\usepackage[a4paper]{geometry}
\usepackage{amssymb}
\usepackage{amsmath}
\usepackage{textcomp}
\usepackage{array}	
\newcolumntype{M}{>{$\vcenter\bgroup\hbox\bgroup}c<{\egroup\egroup$}}
\usepackage{verbatim}
\usepackage{ulem}
\usepackage[caption=false]{subfig}

\usepackage[pdftex]{graphicx}
\usepackage[pdftex]{hyperref}

\usepackage{graphicx}     
\DeclareGraphicsExtensions{.png, .jpg, .pdf, .tif}

\usepackage[usenames,dvipsnames]{xcolor}
\definecolor{MyGreen}{RGB}{0,200,100}
\graphicspath{{.},{Figs/}} 

\hypersetup{
linkcolor=RubineRed,     
citecolor=blue,     
urlcolor=blue,      
colorlinks=true,
citebordercolor=0 0 0,  
}

\newcommand{\tper}{\tau_\textrm{per}}
\newcommand{\epss}{\epsilon_\textrm{s}}
\newcommand{\mylab}[1]{\label{#1}} 		

\renewcommand{\vec}[1]{\mathbf{#1}}

\begin{document}

\title{Directed transport of polymer drops on vibrating superhydrophobic substrates: A Molecular Dynamics study}
\author {Nikita Tretyakov}
\email{tretyakov@mpip-mainz.mpg.de}
\affiliation{Max-Planck-Institut f\"{u}r Polymerforschung, Ackermannweg 10,
55128 Mainz, Germany}
\affiliation{Institut f\"{u}r Theoretische Physik, Georg-August-Universit\"at, Friedrich-Hund-Platz 1,
37077 G\"{o}ttingen, Germany}
\author {Marcus M\"{u}ller}
\email{mmueller@theorie.physik.uni-goettingen.de}
\affiliation{Institut f\"{u}r Theoretische Physik, Georg-August-Universit\"at, Friedrich-Hund-Platz 1,
37077 G\"{o}ttingen, Germany}

\begin{abstract}
Using Molecular Dynamics simulations of a coarse-grained polymer liquid we investigate the transport of droplets on asymmetrically structured (saw-tooth shaped), vibrating substrates. Due to a continuous supply of power by substrate vibrations and the asymmetry of its topography, the droplets are driven in a preferred direction. We study this directed motion as a function of the size of the droplets, the linear dimensions of the substrate corrugation, and the period of vibrations. 
  
Two mechanisms of driven transport are identified: (i) one that relies on the droplet's contact lines and (ii), in a range of vibration periods, the entire contact area contributes to the driving.  In this latter regime, the set-up may be used in experiments for sorting droplets according to their size. Additionally, we show that the linear dimension of the substrate corrugation affects the flux inside the droplet. While on a substrate with a fine corrugation droplets mostly slide, on a more coarsely corrugated substrate the flux may exhibit an additional rotation pattern.
\end{abstract}

\maketitle

\section{Introduction}
\mylab{sec:intro}

An ability to drive a liquid droplet in a controlled way is crucial for micro- and nanofluidics. For instance, directed motion could be used to transport substances suspended in the drop to distinct parts of the lab-on-a-chip device. It will also provide an opportunity to sort droplets of a specific size needed in some applications if the transport mechanism exhibits a size dependence.

In order to actuate a droplet sitting on a substrate, one has to employ surface-energy gradients. These gradients may be divided into two general groups~\cite{AS_MC_KB_2006}: static and dynamic. Prototypical examples of the former ones are methods based on asymmetric dewetting, i.e., solids patterned by posts of variable density~\cite{AS_MC_KB_2006,MR_FP_DQ_2009} or chemical gradients~\cite{MC_GW_1992}. Dynamic gradients, on the other hand, are made by changing the wettability in time, for example, by electrowetting~\cite{FM_LB_2005} or continuous reactive wetting~\cite{FD_TO_1995,SWL_DK_PL_2002}. Some actuating approaches use a combination of both, static and dynamic, gradients. The drop on a static substrate can be driven by dynamical temperature gradients~\cite{Brochard_1989}, driving body forces,~\cite{HK_JL_AD_JY_2006, JS_MM_2008} or vibrations of the solid substrate~\cite{SD_MC_2002}. 
The vibrations in the latter case can be either asymmetric on symmetric substrates or symmetric on asymmetric substrates~\cite{AB_LT_PS_2002,KJ_UT_2007,KJ_UT_2010}.

In this paper we study the directed transport of droplets on asymmetrically structured, vibrating substrates (ASVS). The response of the drops is studied as a function of its size, linear size of a substrate corrugation, and period of vibrations. Using Molecular Dynamics simulations we aim at answering three questions: 
\begin{itemize}
 \item {\it What is the driving mechanism?} For instance, the asymmetric hysteresis of the advancing and receding contact lines (CLs)~\cite{AB_LT_PS_2002,MP_DQ_JB_2008} might drive the directed transport. Alternatively, the fluid flow at the vibrating polymer-substrate area could set the drop in motion. Since both driving mechanisms differ in their dimension -- three-phase contact line versus contact area -- a systematic study of droplet size effects can distinguish them. 
 \item {\it What is the character of droplet motion?} In general, there are two limiting possibilities~\cite{BM_HK_JY_2010,JS_MM_2008}: sliding or rolling motions. The character of the fluid flow inside the droplet dictates how the energy input is dissipated and it may depend on the period of substrate vibrations, $\tau_\textrm{per}$, the strength of solid-liquid interaction, $\epsilon_\textrm{s}$, and the linear size of the substrate's corrugation. 
 \item {\it What is the efficiency of directed motion?} To answer this question we examine dissipation mechanisms for drops of varying size and relate them to the power input due to the substrate vibrations. 
\end{itemize}

Our manuscript is organized as follows: In the next section~\ref{sec:model}, we present the topography of the substrate, the polymer model and simulation techniques. Then, in section~\ref{sec:agit:response}, we explain the response of a droplet to agitation and the direction of motion. The mechanisms of directed motion are reported in section~\ref{sec:agit:CAdrive}. The subsequent section~\ref{sec:agit:mech} deals with the character of directed motion, its efficiency and dissipation mechanisms. We conclude with a short discussion in section~\ref{sec:agit:summ}.

\section{Model}
\mylab{sec:model}

Here, we describe the  model of the polymer liquid and the topography of ASVSs and provide details about the simulation techniques.

The liquid droplet consists of short coarse-grained polymer chains that are comprised of $N_\textrm{p}=10$ segments. By virtue of the vanishingly small vapor density,  $\rho_{\rm{V}}/\rho_{0} \sim 10^{-5}$, relative to that of the polymer melt, $\rho_{0}$, the vapor pressure is so low that evaporation effects are negligible.

Monomers along a flexible macromolecule are bonded by a finitely extensible nonlinear elastic (FENE) potential \cite{Bird77, KK_GG_1990}. Additionally, short-range excluded volume and longer-ranged attractions inside the fluid are represented by a shifted 12-6 Lennard-Jones (LJ) potential {$U(r)=U_\textrm{{LJ}}(r)-U_\textrm{LJ}(r_\textrm{cut})$, where}
\begin{equation}
	U_\textrm{LJ}(r)=4 \epsilon \bigg[ \Big( \frac{\sigma}{r}\Big)^{12}-\Big( \frac{\sigma}{r} \Big)^6 \bigg].
\mylab{eq:LJ}
\end{equation} 
$r$ denotes the distance between segments, and $\epsilon$ and $\sigma$ characterize the energy and length scales, respectively. The non-bonded interactions between segments are cut off at $r_\textrm{cut}=2\times 2^{1/6} \sigma$. 

The substrate is a LJ crystal constructed of face-centered-orthorhombic ({\it fco}) unit cells with number density $\rho_\textrm{s} = 2.67 \sigma^{-3}$. The lattice vector lengths are $a_x = a_y = \sqrt{3} a$ and $a_z = a$, where $a = \sqrt[3]{0.5}\sigma$. The particles of the substrate interact with the polymer segments by a LJ potential $U_{s}$ which has a similar form as Eq.~(\ref{eq:LJ}). 
The length scale of this polymer-substrate interaction is $\sigma_\textrm{s}=0.75 \sigma$, and we employ $\epss = 0.4\epsilon$. On a flat substrate, this polymer-substrate interaction results in the equilibrium contact angle $\Theta_{\rm e} = 138.1\text{\textdegree}$ and the bulk number density of the liquid is $\rho_0 = 0.786 \sigma^{-3}$~\cite{NT_MM_slip_12} at liquid-vapor coexistence at temperature $k_\textrm{B}T = 1.2\epsilon$.

The equations of motion are integrated with the velocity-Verlet algorithm using an integration time step of $\Delta t= 0.005 \tau$ in LJ time units $\tau=\sigma \sqrt{\frac m\epsilon}$ (with $m$ being the segment mass). Temperature is controlled by a Dissipative Particle Dynamics (DPD) thermostat \cite{PH_JK_1992,PE_PW_1995}. Since it provides local momentum conservation inside the liquid, hydrodynamic behavior is observed on large time and length scales. However, in contrast to soft substrates~\cite{FL_JS_CP_MM_2011} or an Einstein crystal substrate, we keep the atoms of the substrate frozen at the nodes of the crystal lattice. Therefore, the polymer-substrate interactions break local momentum conservation and the center of mass of the polymer droplet can spontaneously change its velocity. 

\begin{figure}[t]
    \subfloat[]{\label{fig:drop_s3d0_n19712}\includegraphics[width=0.45\hsize]{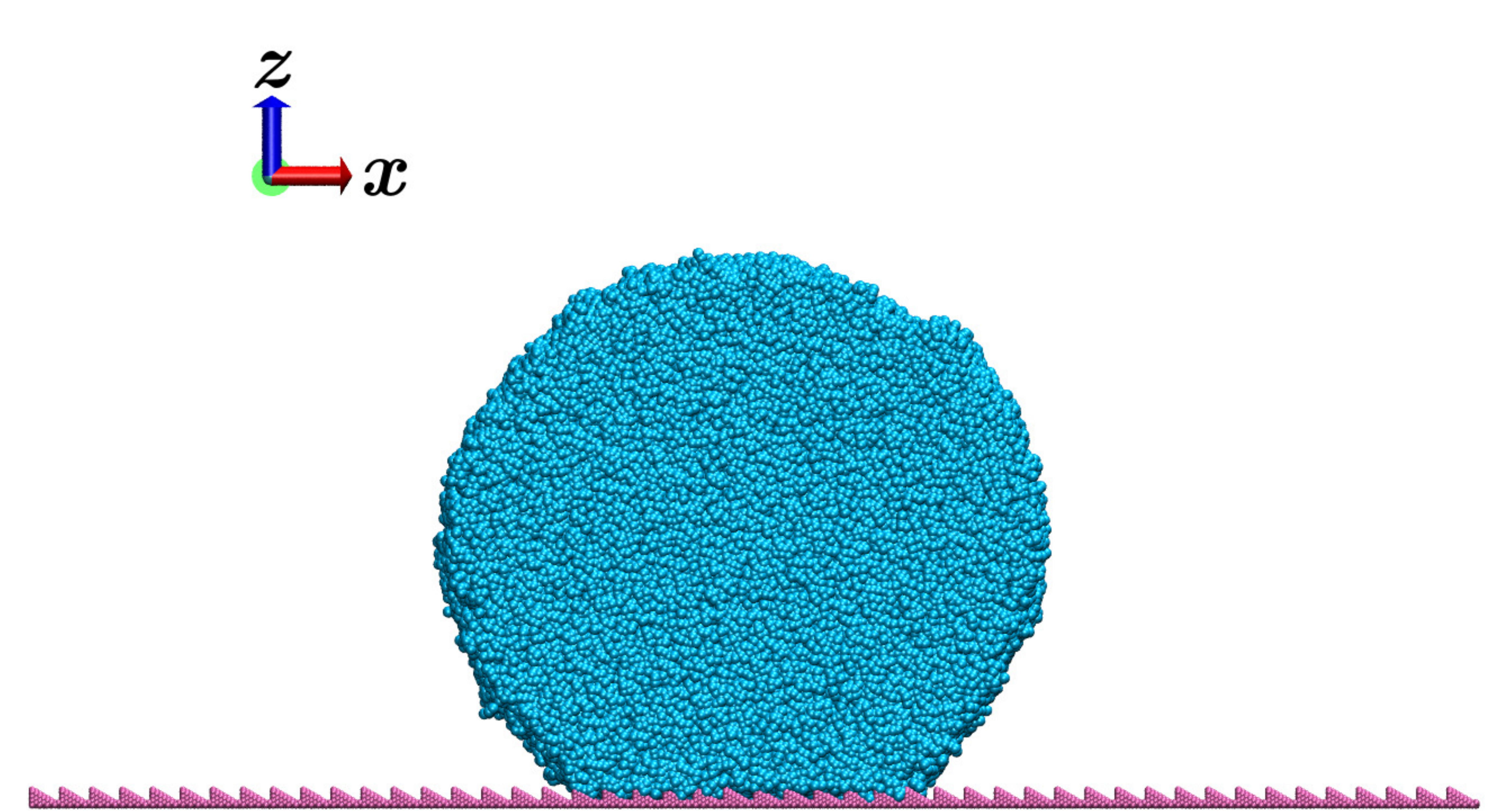}}\hspace{0.5cm}
    \subfloat[]{\label{fig:sub_s3d0_s7d0}\includegraphics[width=0.45\hsize]{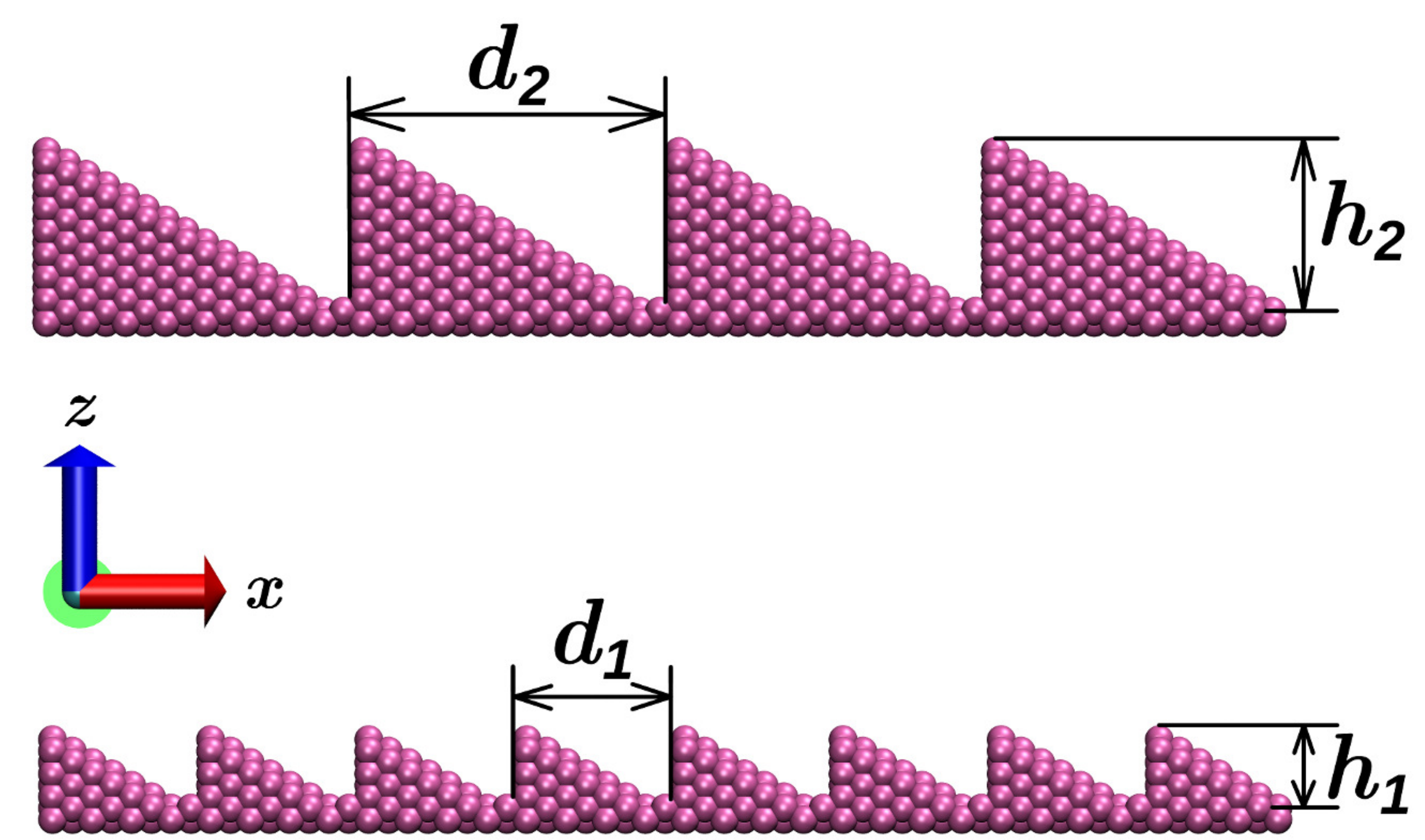}}
 \caption{(a) A cylindrical liquid drop on an asymmetrically structured substrate. Side view. (b) The geometries of asymmetrically structured substrates. Two substrates with grooves of different size are studied: roughly corrugated, R (top) and finely corrugated, F (bottom). The dimensions of the former one are $d_1 = 4.81\sigma$ and $h_1=2.78\sigma$. The latter one is twice as large, $d_2 = 9.62\sigma$ and $h_2=5.56\sigma$. The angle of corrugation is $30\text{\textdegree}$ in both cases.}
 \label{fig:system_substrates}
\end{figure}

The polymer droplet is contained in a three-dimensional simulation domain of volume $V=L_x \times L_y \times L_z$. Periodic boundary conditions are used in $x$- and $y$-directions, whereas an ideal repulsive wall is placed far above the droplet, sitting on a supporting substrate, in $z$-direction. The geometry {of the} simulation box,  $L_y=30.25\sigma \ll L_x$, is chosen such that the droplet spans the neutral $y$-direction forming a cylindrical droplet or ridge. The side view of a system is shown in Fig~\ref{fig:drop_s3d0_n19712}, the axis of a cylinder is parallel to the $y$-axis of the box. We note that due to the cylindrical geometry, one may (i) study drops of bigger radii than in the case of spherical droplets~\cite{JS_MM_2008} and (ii) neglect line tension effects,~\cite{Indekeu94, NT_MM_DT_UT_2012} for the length of the three-phase contact lines, $2L_{y}$, does not depend on the contact angle.

We study substrates with fine (F-type) and rough (R-type) corrugations. The grooves of both substrates have triangular cross-section as depicted in Fig.~\ref{fig:sub_s3d0_s7d0}, but differ in size by a factor of $2$. The left side wall {of the corrugation} is vertical while the right one forms a $30\text{\textdegree}$ angle with the substrate plane. This substrate {topography} in conjunction with the high contact angle on a planar substrate {at} the same polymer-substrate interaction strength, $\epss$, leads to the superhydrophobicity.

In order to set the drop in motion we vertically oscillate all substrate atoms in-phase, such that the mutual distances between substrate particles remain unaltered. Their velocity is given
\begin{equation}
 \vec{v^{\rm{s}}}(t) = v_z^{\rm{s}}(t) \vec{n}_{z} = A \vec{n}_{z} \, \omega \sin (\omega [t - t_0]),
 \label{eq:vib_law}
\end{equation}
where $A=1\sigma$ is the amplitude of the vibrations chosen to be small even with respect to the radius of the smallest droplet ({\it ca.} $17\sigma$). $\omega$ and $t_0$ denote the angular frequency and the time when agitation starts, respectively. $\vec{n}_{z}$ is the unit vector normal to the $xy$-plane. The initial velocity $v_z^{\rm{s}}(t_0)$ is set to zero in order to avoid an initial energy or momentum jump. This facilitates the equilibration towards a steady state and prevents the detachment of the drops from the substrate at the start of the vibrations.

\begin{figure}[t]
	\subfloat[]{\label{fig:prof_s3d0_e04_41}\includegraphics[width=0.45\hsize]{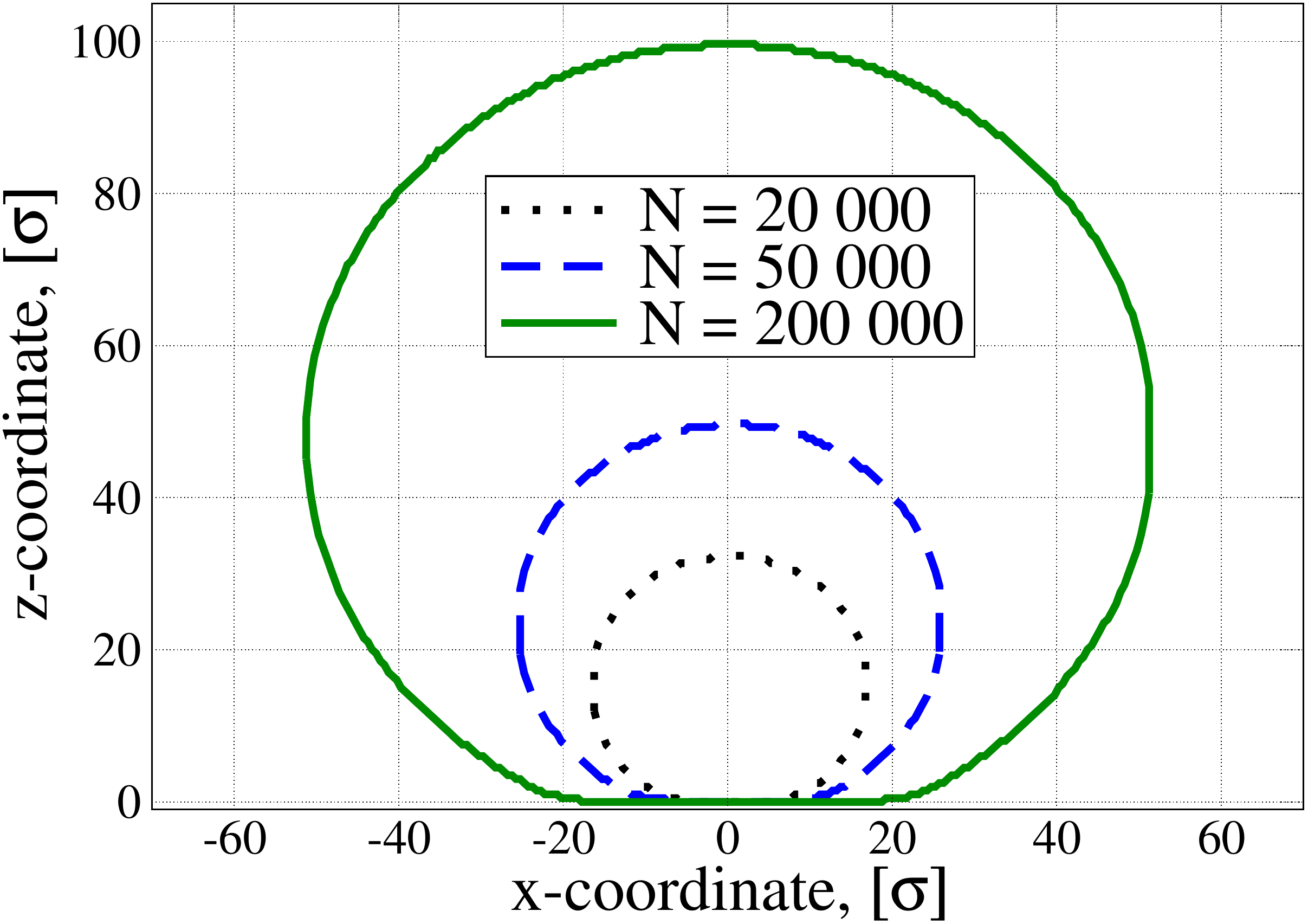}}\hspace{0.2cm}
	\subfloat[]{\label{fig:prof_s3d0_n19712}\includegraphics[width=0.45\hsize]{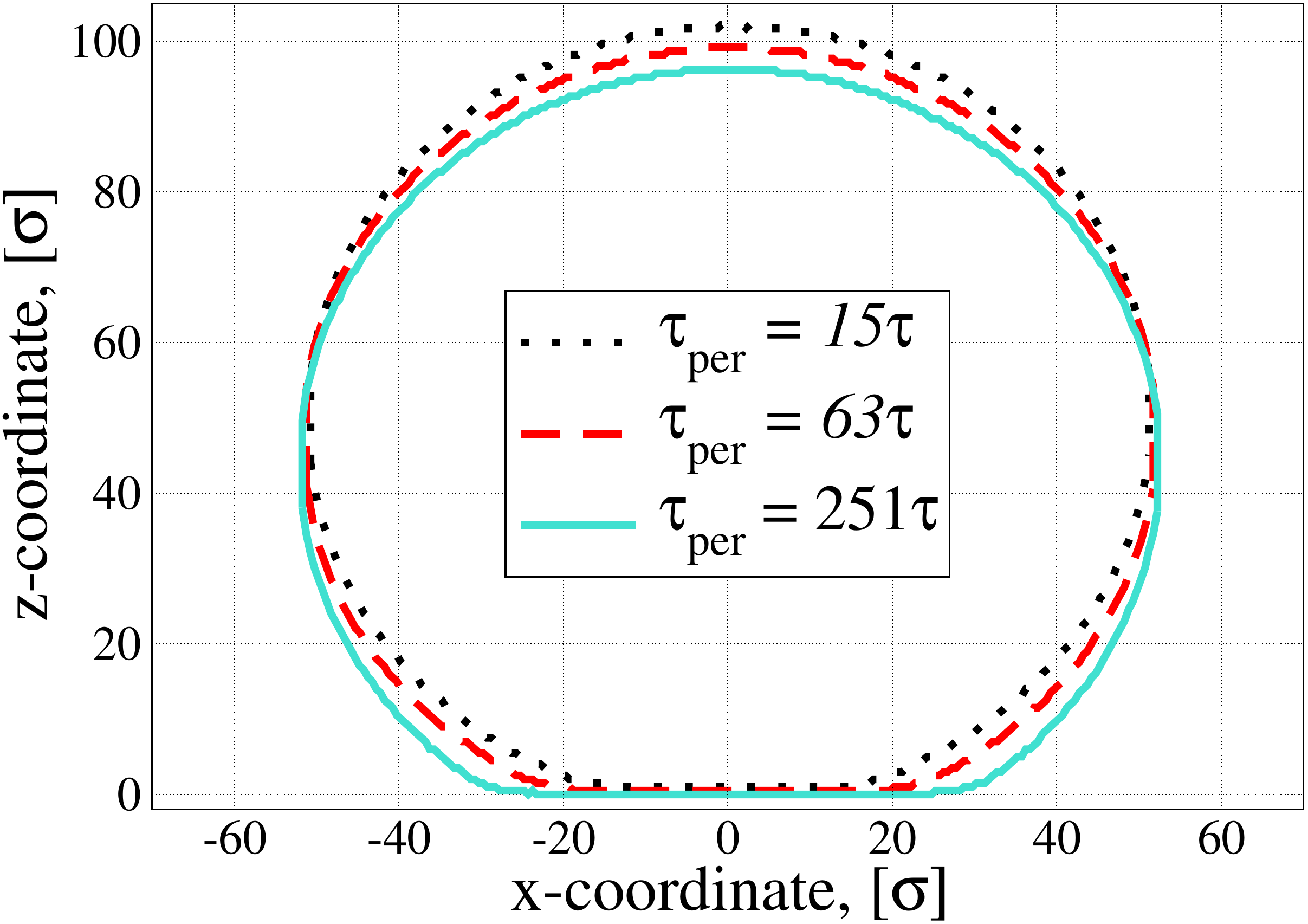}}
	\caption{(a) Profiles of the drops of various size on the F-type substrate at the period of substrate vibration $\tau_\textrm{per}=41\tau$. (b) Profiles of drops comprised of $N=200\,000$ beads on the F-type substrate. Different line types represent profiles at various periods of substrate oscillation.}
	\label{fig:pos_prof}
\end{figure}

The response of the droplets to the substrate vibration is investigated for vibration periods in the interval $15 \tau \leqslant \tau_\textrm{per} = 2\pi/\omega \leqslant 251 \tau$. This interval includes the Rouse relaxation time, estimated for the polymer liquid to be $\tau_\textrm{R} = 25.3 \pm 5\,\tau$~\cite{NT_MM_slip_12}. To quantify the dependence of the mechanism and the efficiency of directed motion on the size of the droplet, we investigate drops comprised of {{\it ca.}} $20\,000$, $50\,000$ or $200\,000$ beads. Their profiles at $\tau_\textrm{per}=41\tau$ on the F-type substrate are displayed in Fig.~\ref{fig:prof_s3d0_e04_41}. Note, that not only the radii of the droplets differ, but also the contact areas (CA). The contact angles, however, remain equal within a standard deviation but they depend on the vibration period. The later is illustrated in Fig.~\ref{fig:prof_s3d0_n19712} and will be discussed in the next section.

In the following we present results for the steady state of running droplets. For a better statistics we performed at least $15$ independent runs for each parameter set  specified by droplet size $N$, period of vibrations $\tau_\textrm{per}$,  and substrate type. All of them started from the independent initial configurations. Every run extended to $2\times10^7$ steps or, equivalently, $10^5 \tau$. We extensively used graphical processing units (GPU) in conjunction with the HOOMD-blue simulation package ~\cite{HOOMD_site, JA_CL_AT_2008, CP_JA_SG_2011} modified to describe the substrate vibrations.

\section{Direction of motion}
\mylab{sec:agit:response}

\begin{figure}[t]
	\subfloat[]{\label{fig:explain_dir_0}\includegraphics[width=0.3\hsize]{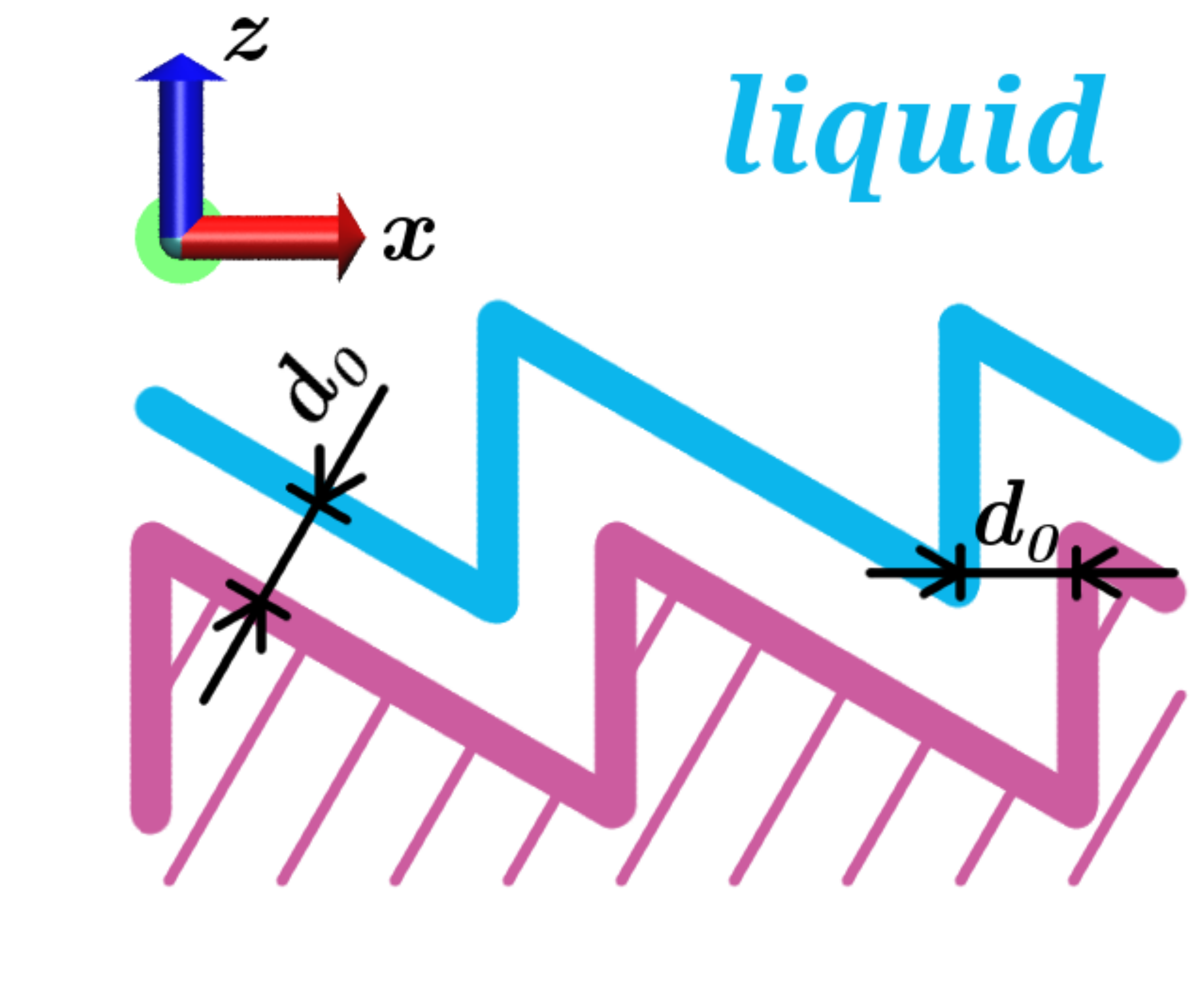}}\hspace{0.5cm}
	\subfloat[]{\label{fig:explain_dir_up}\includegraphics[width=0.3\hsize]{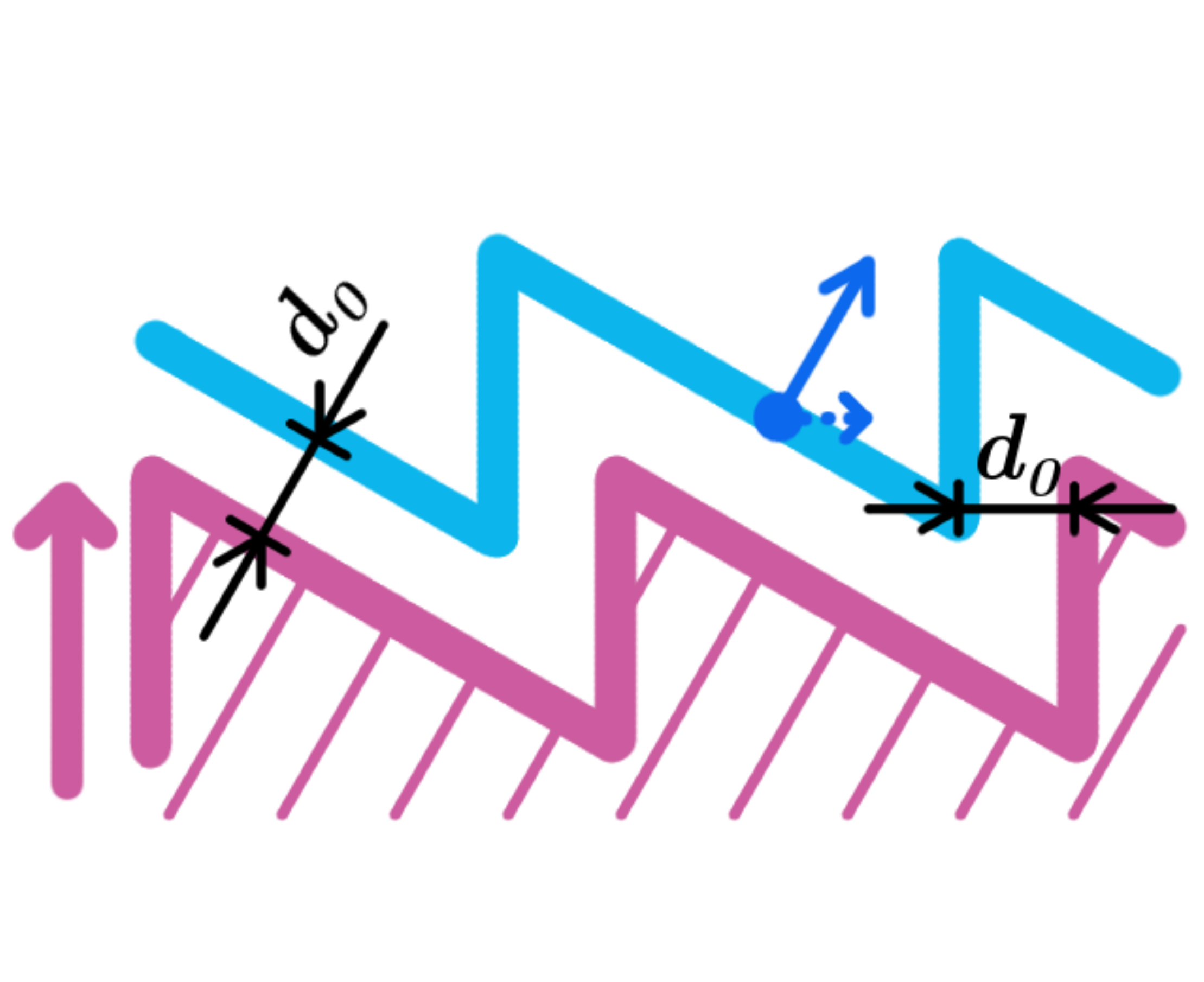}}\hspace{0.5cm}
	\subfloat[]{\label{fig:explain_dir_down}\includegraphics[width=0.3\hsize]{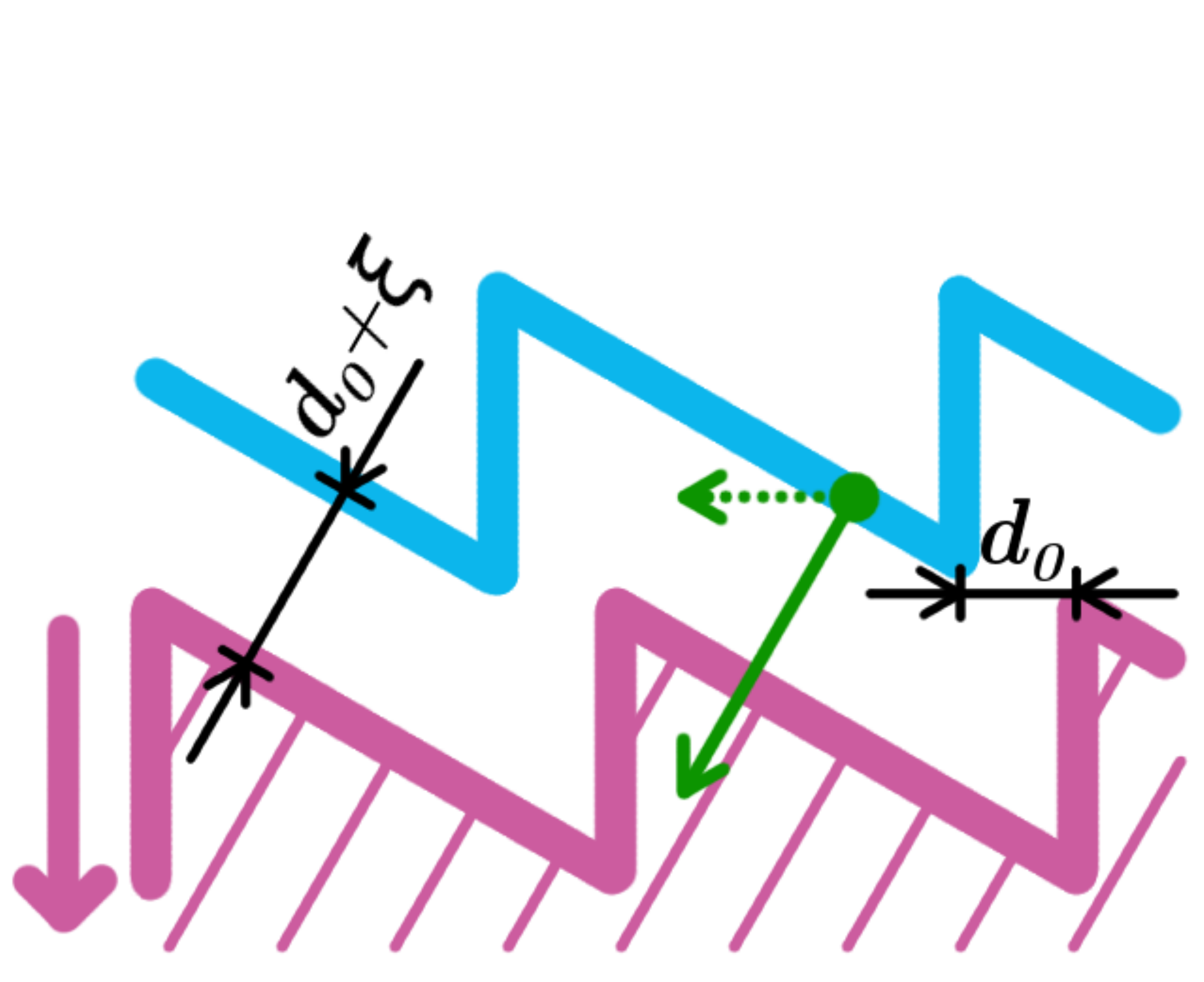}}
	\caption{Explanation of the direction of motion on the ASVS (of F-type): (a) The solid-liquid interface (blue contour) of the droplet resting on a asymmetrically structured substrate (lila contour). The distance from the interface to the solid is $d_0$. (b) {Upward motion} of the ASVS is marked by the lila arrow. The liquid is pushed by the solid (blue arrow), and the repulsion drives the droplet to the right (blue pointed arrow).
	(c) {Downward motion} of the ASVS. The distance from the solid changes by amount of $\xi$, shifting the effective interaction into attractive range of the potential (green arrow). It has a component directed to the left (green pointed arrow) that on average is greater than repulsive one and provides the directed motion to the left.}
	\label{fig:explain_dir}
\end{figure}

In this section we qualitatively explain the directed motion by simple phenomenological considerations.
A droplet placed on an immobile asymmetrically structured substrate, as shown in Fig.~\ref{fig:explain_dir_0}, can not move in any specific direction. Taking advantage of thermal fluctuations, it would act like a Maxwell's demon \cite{Maxwell_1871} and a directed motion would violate the second law of thermodynamics. However, if one constantly supplies energy to the system, e.g.~by vibration of the substrate, the droplet can move in a preferred direction via a ratchet mechanism~\cite{Magnasco_93,Hanggi96,Julicher97,Reimann02}. Two driving forces contribute to the directed transport of the droplet.

(i) {\it Contact-line driving}: The pinning of the CL at the ridges of the substrates gives rise to contact angle hysteresis. According to the Concus-Finn criterion~\cite{ConcusFinn} the contact angle, $\Theta_{\rm left}$ (measured with respect to the horizontal) at the right CL can adopt stable values in the interval  $\Theta_{\rm e} - 90\text{\textdegree} \leq \Theta_{\rm right}  \leq \Theta_{\rm e} + 30\text{\textdegree}$ with $ \Theta_{\rm e} = 138.1\text{\textdegree}$. If $\Theta_{\rm right}$ reaches the lower bound, the right CL will move down the vertical slope (to the left); if it reaches the upper bound, the right CL will slide down the gently inclined slope to the right. 

For the left CL the condition reads: $ \Theta_{\rm e}-30\text{\textdegree}\leq \Theta_{\rm left} \leq \max(180\text{\textdegree},\Theta_{\rm e}+90\text{\textdegree})$. At the lower bound the left CL will move to the right. The upper bound is modified because the left CL will discontinuously jump to the neighboring ridge on the left hand side (tank-treading) if the contact angle $\Theta_{\rm left}$ exceeds $180\text{\textdegree}$. 

The average contact angle on the vibrating substrate depends on the period of vibrations, $\tau_\textrm{per}$, as shown in Fig.~\ref{fig:prof_s3d0_n19712}. In the limit of large periods, $\tau_\textrm{per}=251\tau$, the contact angle is similar to that of a non-vibrating substrate. This value on the saw-tooth structured substrate will be larger than the contact angle $\Theta_{\rm e}$ on a planar substrate. If macroscopic considerations were accurate, the contact angle on the saw-tooth substrate in equilibrium would be $\Theta_{\rm e}+30\text{\textdegree}$, i.e.~the contact angle is close to the upper stability limit of the right contact line. Upon decreasing the period of vibration we observe that the contact angle increases indicating that the effective average surface tension between liquid and solid increases.

\begin{figure}[t]
      \subfloat[]{\label{fig:prof_phase_s3d0_e04_15}\includegraphics[width=0.45\hsize]{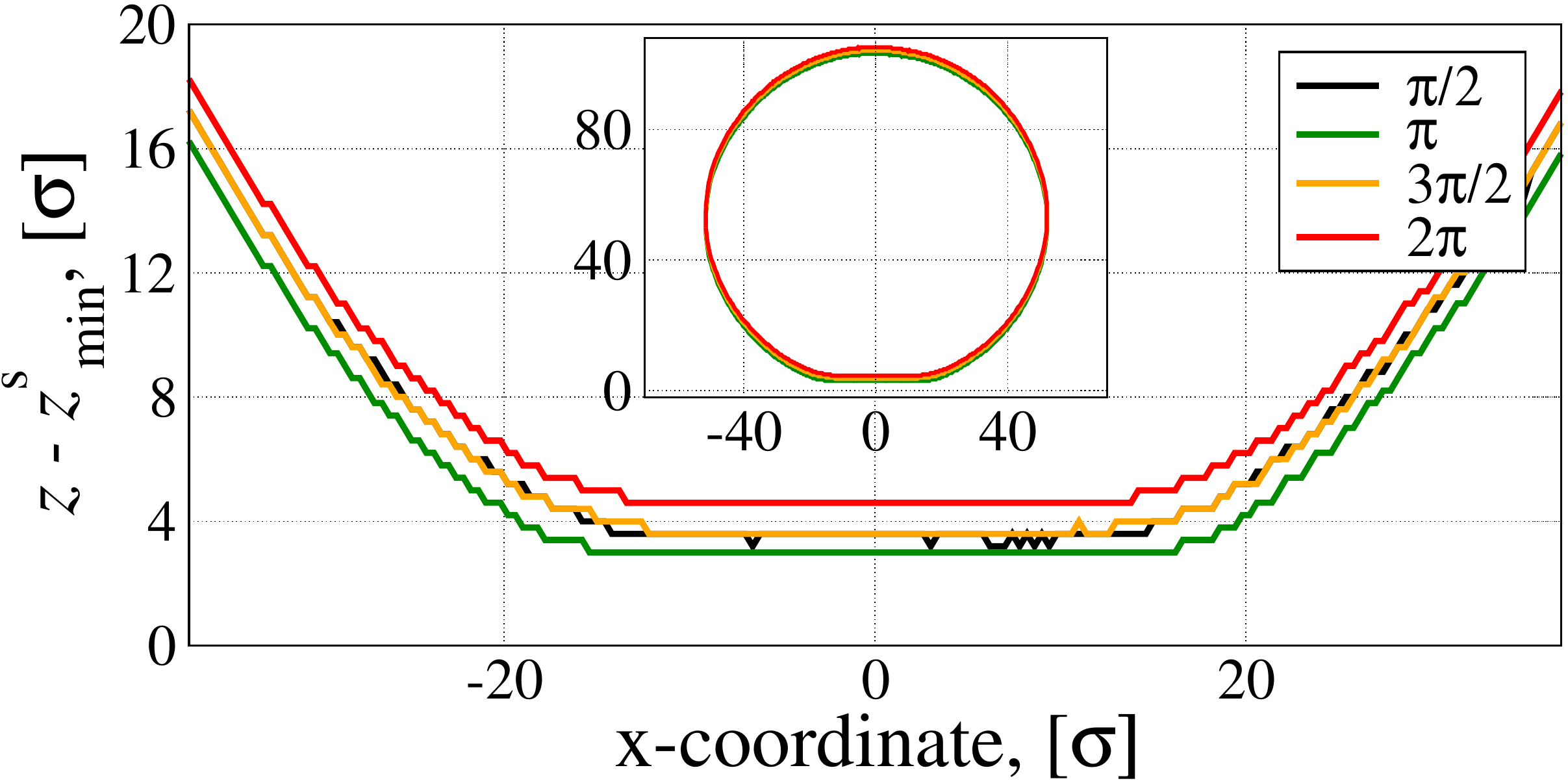}}\hspace{0.5cm}
      \subfloat[]{\label{fig:prof_phase_s3d0_e04_63}\includegraphics[width=0.45\hsize]{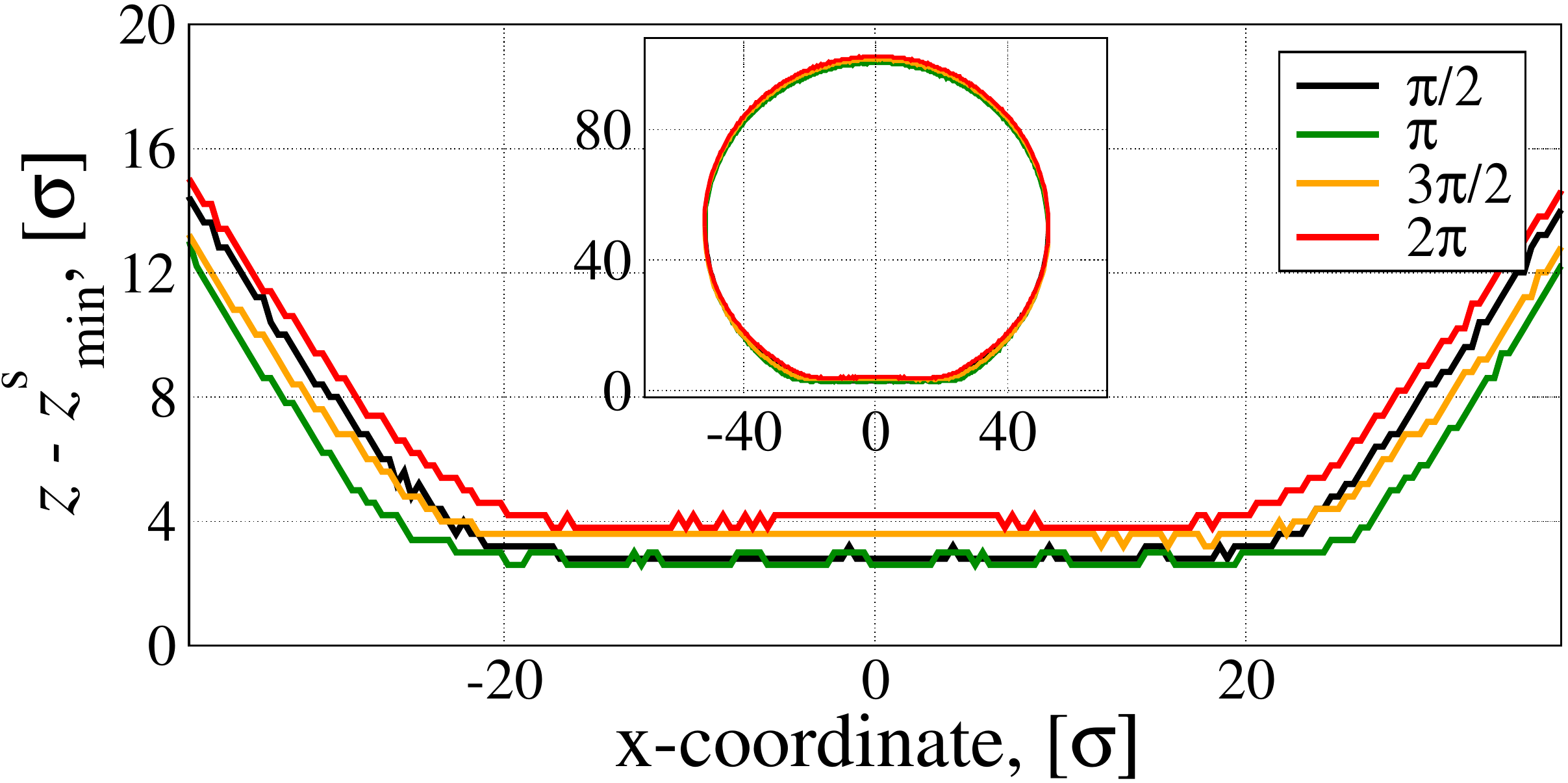}}
      \caption{(a) and (b) show profiles of drops of $N=200\,000$ beads averaged at the same phase $\omega [t - t_0]$ of oscillations on the F-type substrate at $\tau_\textrm{per}=15\tau$ and $63\tau$ vibrating periods, correspondingly. Main panels zoom into the region close to the substrate and show the distance from the solid-liquid interface to the bottom-most layer of the substrate $z^\textrm{s}_\textrm{min}$. The insets show the overall droplet profiles.
      }
      \label{fig:drop_phases}
\end{figure}

During the vibration the distance between the substrate and the polymer liquid is periodically altered. These oscillations give rise to a periodic variation of the instantaneous contact angles because of the flow of the liquid inside of the droplet and inertia effects. If the substrate moves downwards, the instantaneous contact angle increases (cf.~Figs.~\ref{fig:prof_phase_s3d0_e04_15} and~\ref{fig:prof_phase_s3d0_e04_63}) and the right contact line continuously slides down the gently inclined slope. This sliding can be reversed during the upward motion of the substrate. The left CL, however, will jump one substrate periodicity to the left if the contact angle becomes close to $180\text{\textdegree}$ during the downward motion of the substrate, but this discontinuous jump is not reversed during the upward motion. Thus the stick-slip motion of the left contact line drives the droplet to the left.  

We emphasize that these considerations only apply to macroscopic droplets on macroscopically corrugated substrates, where thermal fluctuations of the position of the liquid surface (capillary waves) and the intrinsic width of this surface are negligible compared to the length scale of the substrate, its vibration amplitude and the dimensions of the droplet. The former condition is not met in our simulations because of the microscopic length scale of the substrate. Particularly large deviations must be expected for the F-type substrate, where the length of the jump of the left CL is comparable to the width of the liquid surface and to the length of the downward motion, whereas for the R-type substrate the jump length is $2$ times larger,  $h_1=2.78\sigma$ and $h_2=5.56\sigma$ {\it vs.} $2A = 2\sigma$. 
Therefore, the vapor pockets, which are predicted by the macroscopic considerations, are not developed at the finely corrugated substrate (see Fig.~\ref{fig:explain_dir_0}) and the liquid is in contact with the entire substrate (cf.~the configuration snapshot in Fig.~\ref{fig:liq_under_osc}). Thereby the jump of the left CL is facilitated compared to the roughly corrugated substrate.

(ii) {\it Contact-area driving}:  In addition to the CL hysteresis the entire contact area (CA) between the polymer and the substrate contributes to the directed driving. The direction itself can be justified by the following considerations. 

The liquid is in contact with the substrate on the gently inclined slope. During the upward motion of the substrate the liquid is pushed normal to that slope,  i.e., substrate exerts the force $\vec{F^\textrm{s}}(t)$ (blue solid arrow) onto the liquid as sketched in Fig.~\ref{fig:explain_dir_up}. This time-dependent force has a component $F_{\textrm{up}}(t)\, \vec{i}$ (blue right-directed pointed arrow), where $\vec{i}$ is a unit vector of the $x$-axis. If the solid-liquid interaction were purely repulsive, we would expect the drop to move to the right. In this limit, however, the equilibrium contact angle on a planar substrate would be $180 \text{\textdegree}$ (drying) and the vibration of the substrate would detach the droplet. Instead, we use the LJ potential that comprises an attractive part. Therefore, the opposite half cycle of substrate motion, during which the solid moves downwards as in Fig.~\ref{fig:explain_dir_down}, must also be considered. 
At this stage the solid-liquid distance increases by an amount, $\xi$, and particles of the substrate attract the liquid (green solid arrow). This attraction pulls the liquid to the left with the force $F_\textrm{\,down}(t)\, \vec{i}$ (green pointed arrow). Since the repulsive part of the LJ potential (harsh repulsion) is steep compared to the attractive part, we observe that the average position of the solid-liquid interface is shifted towards larger distances with respect to the minimum of the potential and the integrated force is dominated by the downward cycle resulting in a net force to the left. We also expect this driving mechanism to be stronger on the microscopically corrugated F-type substrate than on the R-type one, because the contact area of the liquid and the substrate is larger for the F-type substrate.

Both mechanisms -- stick-slip motion of CL and CA vibration -- drive the droplet to the left. 

\section{Contact-area driving}
\mylab{sec:agit:CAdrive}

In this section we study further properties of the CA-driven motion. We have averaged the density and velocity profiles in droplets over times at fixed phases $\phi$ of the vibration. From these time-averaged density profiles we locate the position of the liquid surface by requiring that the density is half the density of the bulk liquid (crossing criterion). In Figs~\ref{fig:prof_phase_s3d0_e04_15} and~\ref{fig:prof_phase_s3d0_e04_63} we present the distance of the liquid surface from the substrate for the biggest droplets.  Indeed, the liquid surface is located farther away from the solid when the substrate is moving downwards (phases $3\pi/2$ and $2\pi$, orange and red lines) than when moving upwards (phases $\pi/2$ and $\pi$, black and green lines);
 
\begin{figure}[t]
    \centering
    \subfloat[]{\label{fig:vel_field_s3d0_n19712_e04_15_ph1}\includegraphics[width=0.42\hsize]{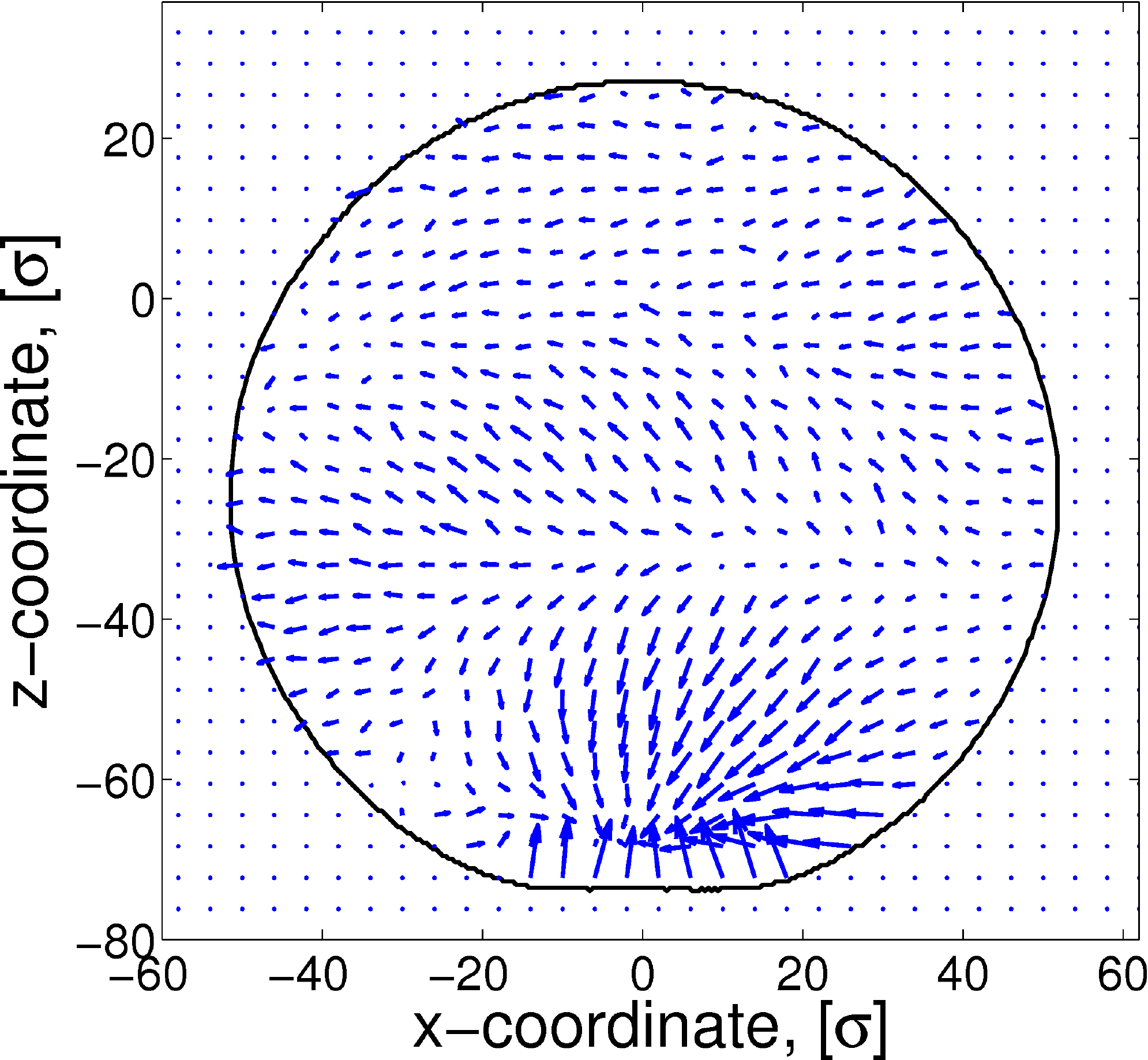}}\hspace{0.1cm}
    \subfloat[]{\label{fig:vel_field_s3d0_n19712_e04_15_ph2}\includegraphics[width=0.42\hsize]{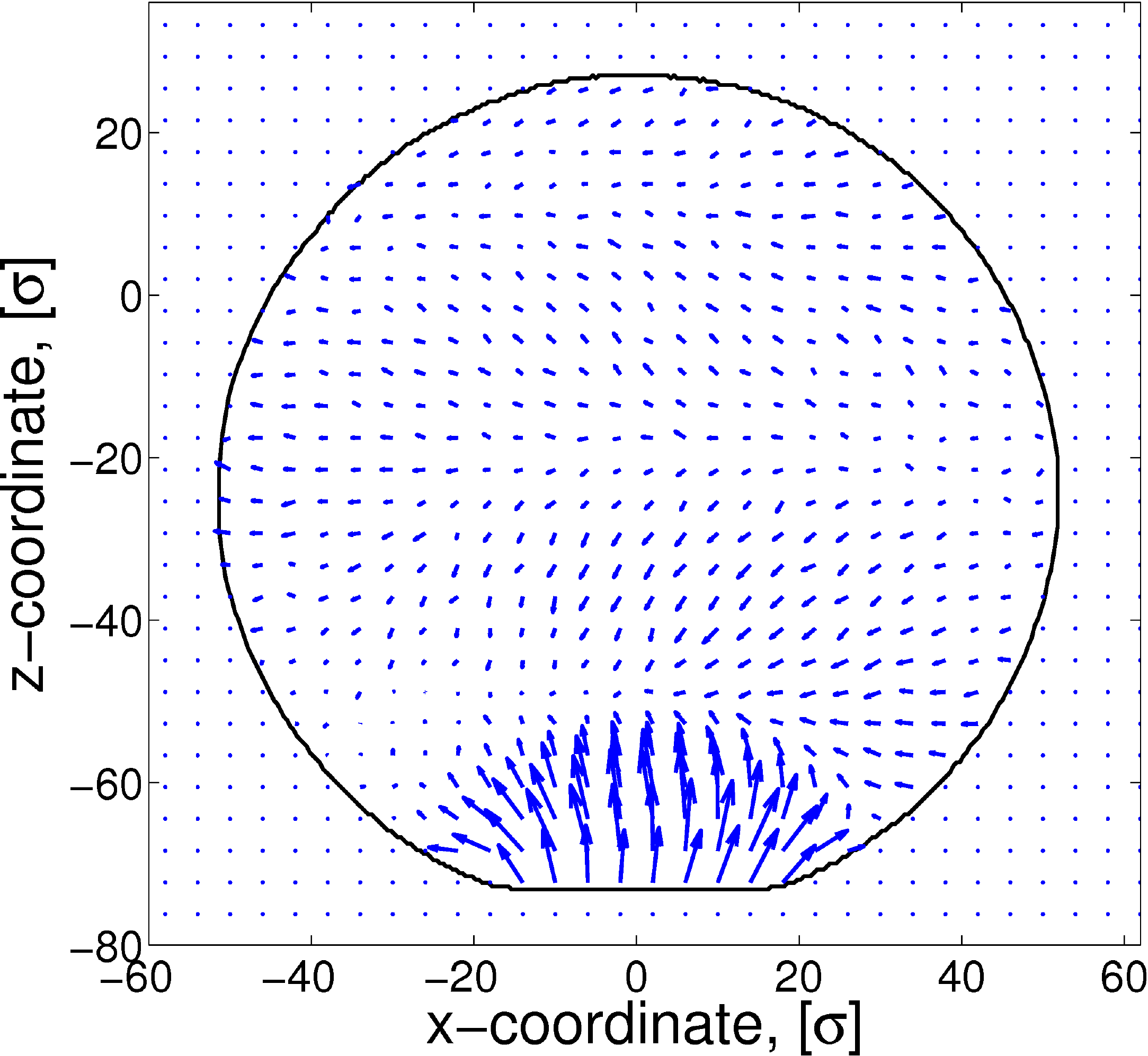}}\hspace{0.1cm}
    
    \subfloat[]{\label{fig:vel_field_s3d0_n19712_e04_15_ph3}\includegraphics[width=0.42\hsize]{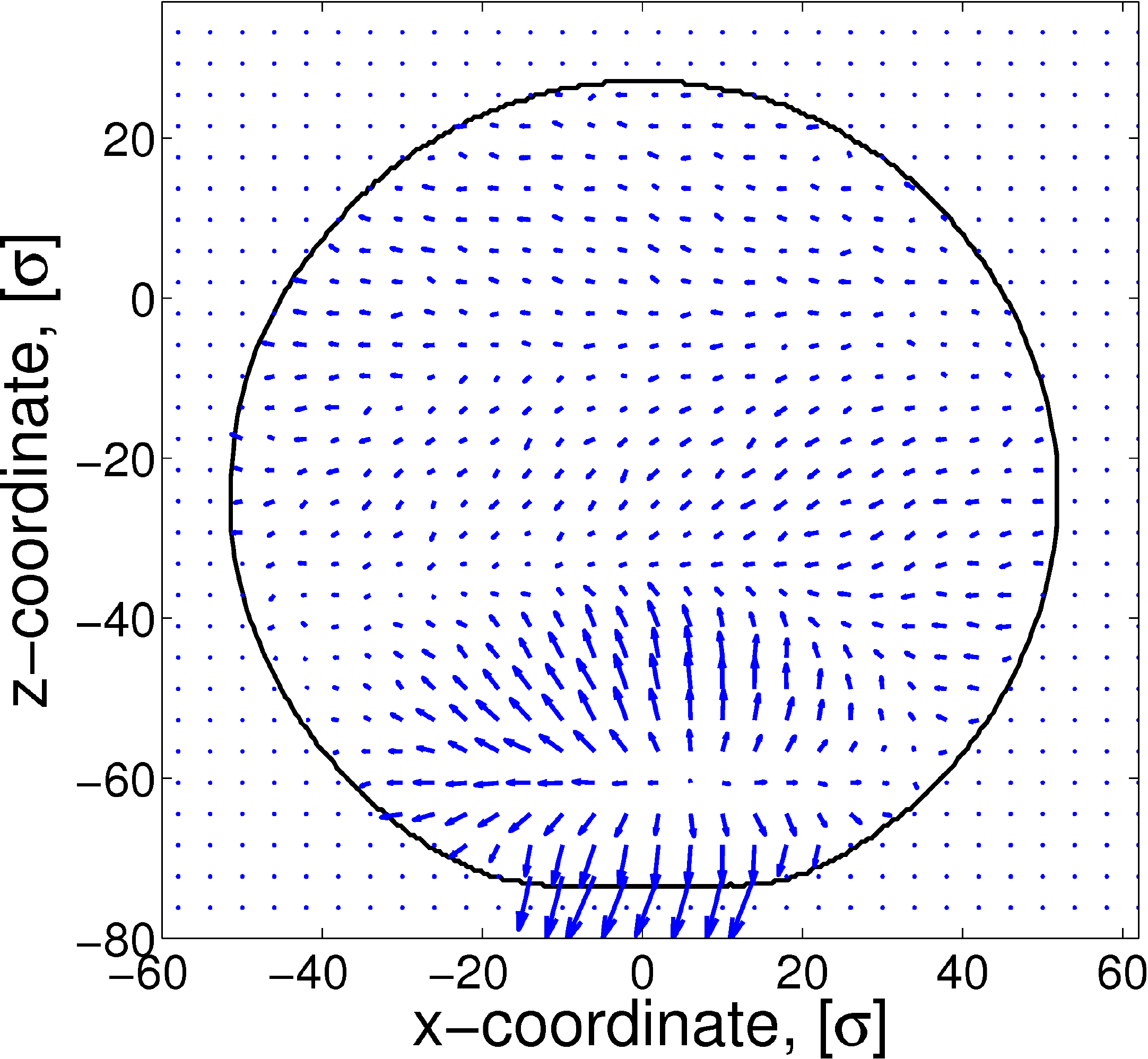}}\hspace{0.1cm}
    \subfloat[]{\label{fig:vel_field_s3d0_n19712_e04_15_ph0}\includegraphics[width=0.42\hsize]{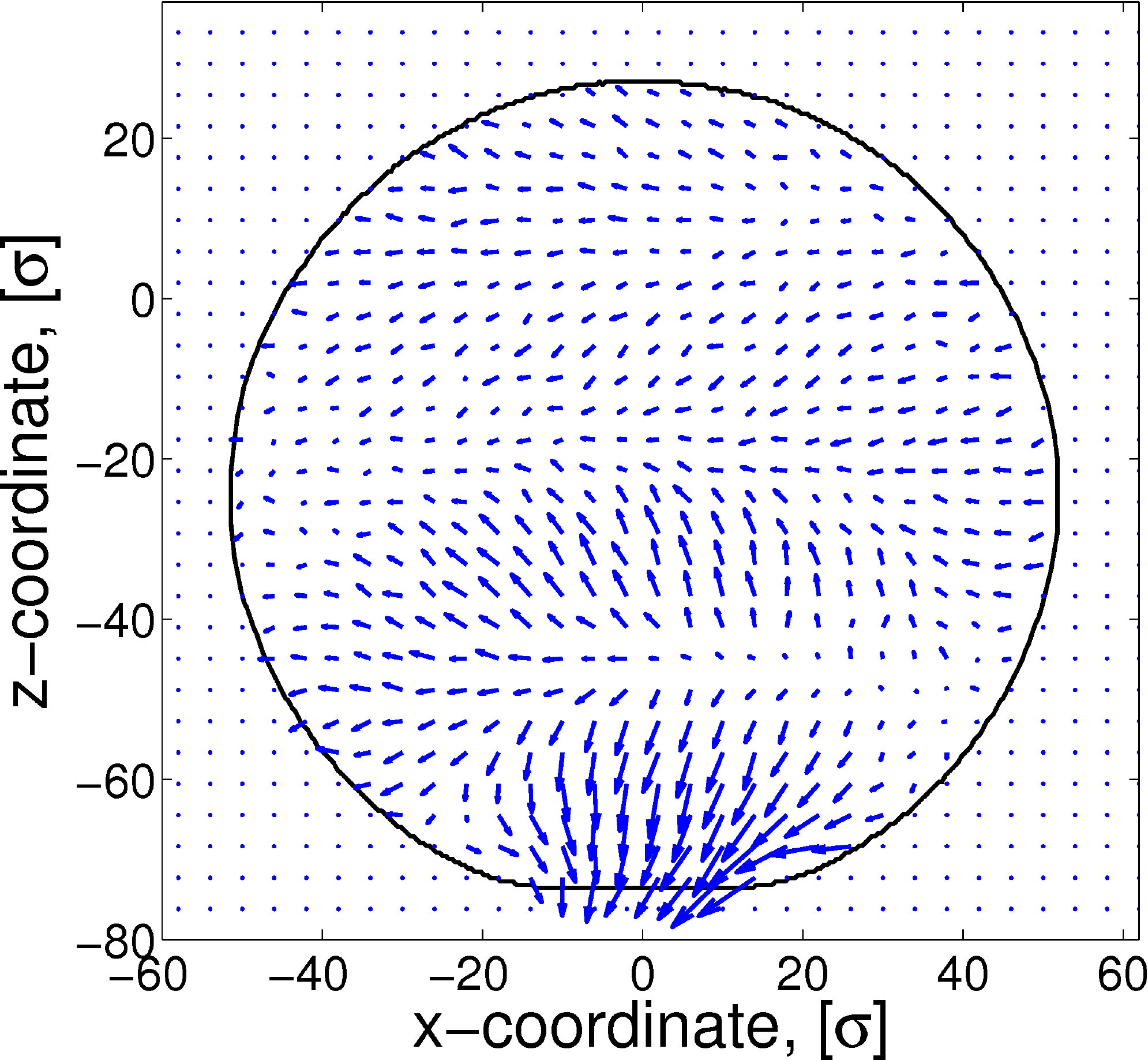}}\hspace{0.1cm}
    \caption{Velocity fields of droplets of $N = 200\, 000$ beads averaged at the same phase $\omega [t - t_0]$ of the F-type substrate vibrations with period of $\tau_\textrm{per}=15\tau$. (a) and (b) correspond to the {upward motion} of the substrate at phases $\pi/2$ and $\pi$, respectively. (c) and (d) correspond to the {downward motion} of the substrate at phases $3\pi/2$ and $2\pi$, respectively. The corrugations of the substrate are not spatially resolved on the profiles at these strength of the solid-liquid interaction and period of vibrations.}
    \label{fig:vel_fields_phases15}
\end{figure}

\begin{figure}[t]
    \centering
    \subfloat[]{\label{fig:vel_field_s3d0_n19712_e04_63_ph1}\includegraphics[width=0.42\hsize]{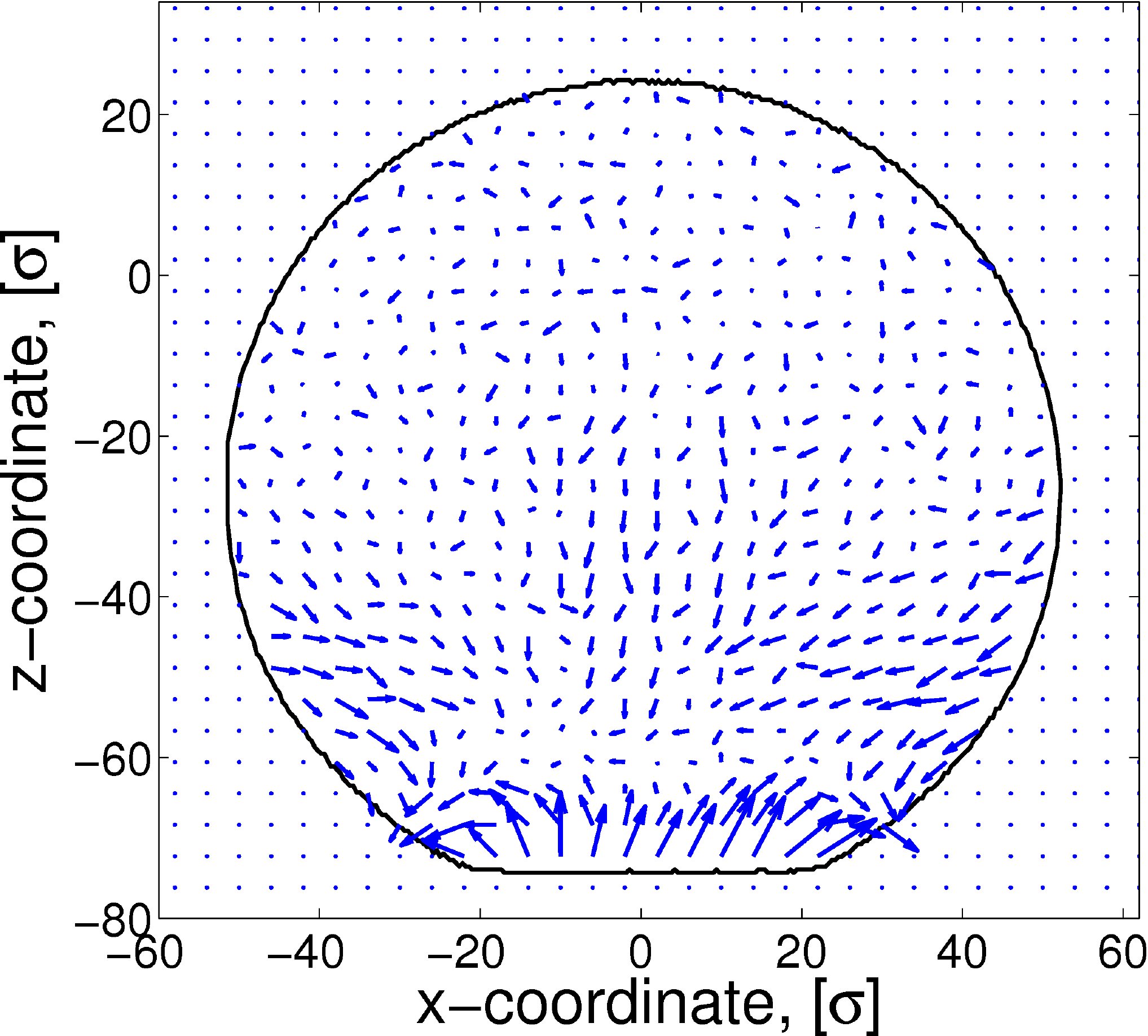}}\hspace{0.1cm}
    \subfloat[]{\label{fig:vel_field_s3d0_n19712_e04_63_ph2}\includegraphics[width=0.42\hsize]{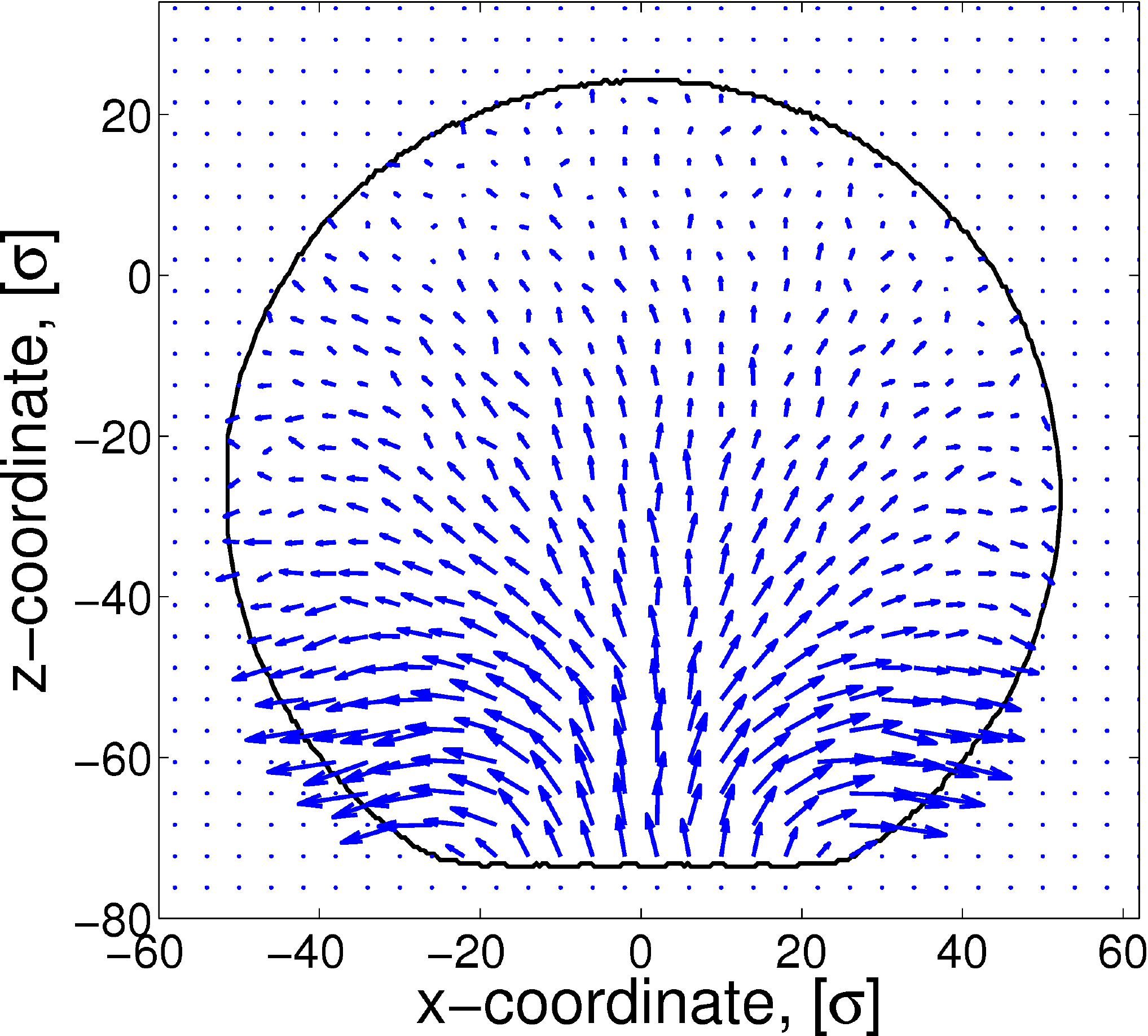}}\hspace{0.1cm}
    
    \subfloat[]{\label{fig:vel_field_s3d0_n19712_e04_63_ph3}\includegraphics[width=0.42\hsize]{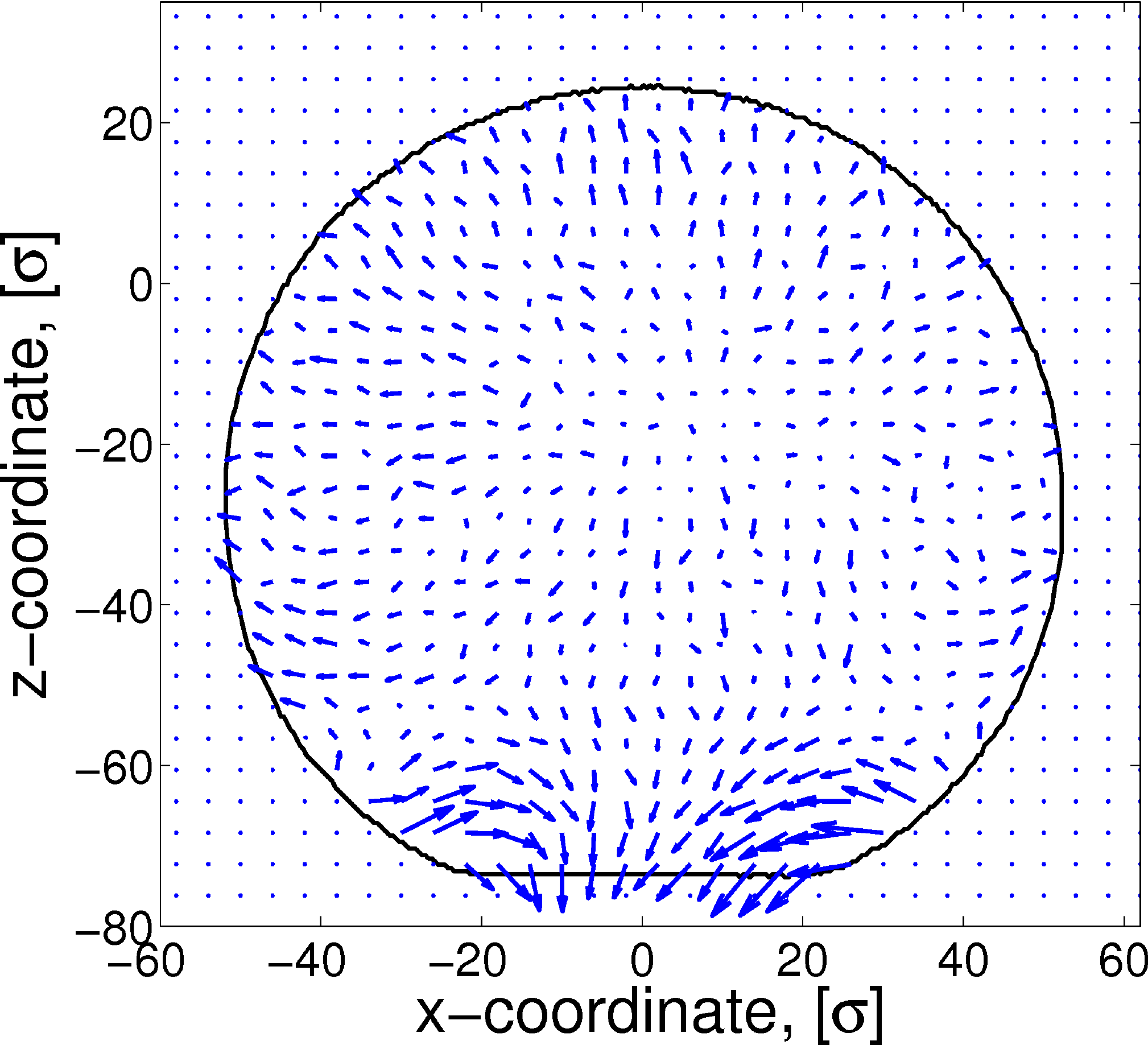}}\hspace{0.1cm}
    \subfloat[]{\label{fig:vel_field_s3d0_n19712_e04_63_ph0}\includegraphics[width=0.42\hsize]{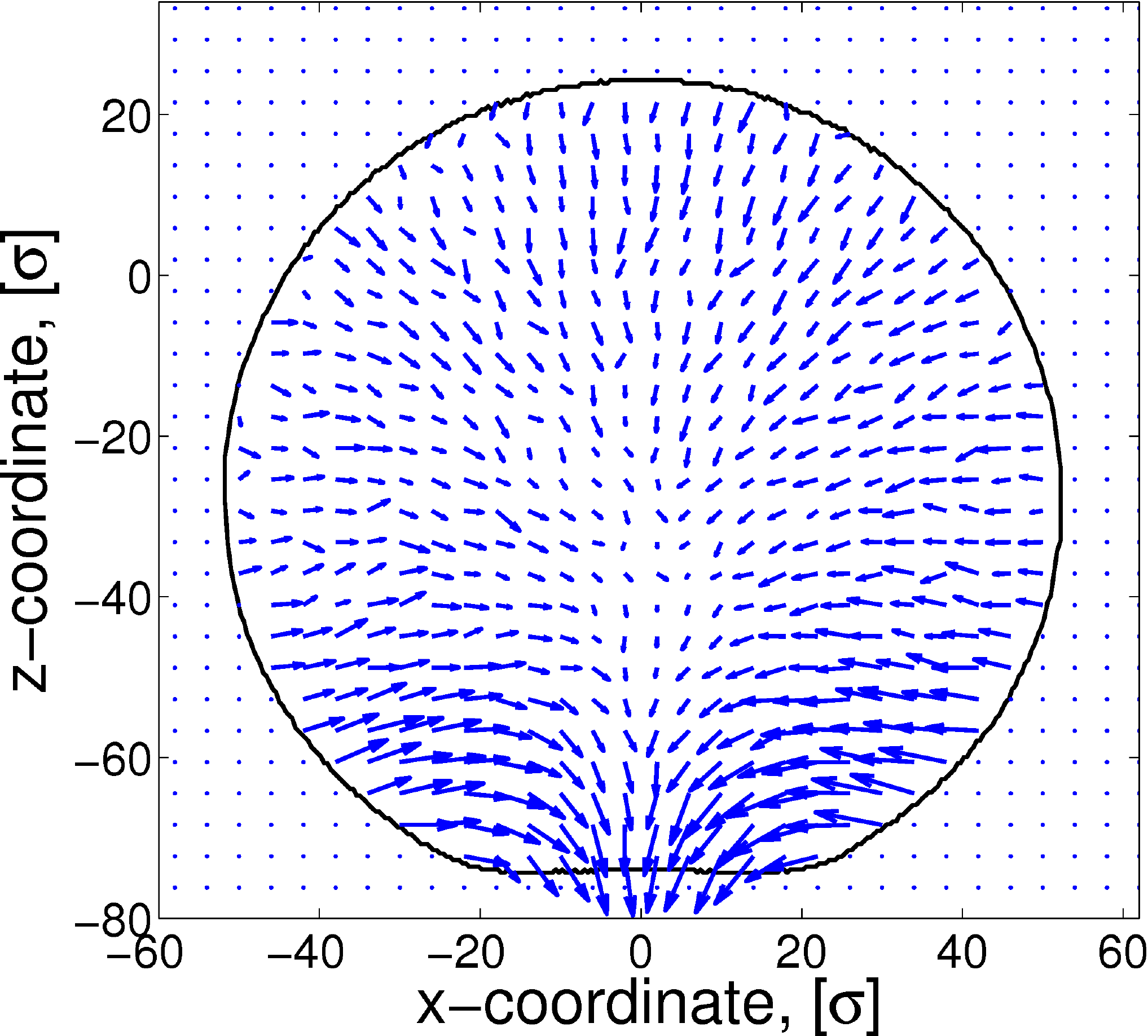}}\hspace{0.1cm}
    \caption{Velocity fields of droplets of $N = 200\, 000$ beads averaged at the same phase $\omega [t - t_0]$ of the F-type substrate vibrations with period of $\tau_\textrm{per}=63\tau$. (a) and (b) correspond to the {upward motion} of the substrate at phases $\pi/2$ and $\pi$, respectively. (c) and (d) correspond to the {downward motion} of the substrate at phases $3\pi/2$ and $2\pi$, respectively.}
    \label{fig:vel_fields_phases63}
\end{figure}

The velocity fields are presented in Figs.~\ref{fig:vel_fields_phases15} and~\ref{fig:vel_fields_phases63} for $\tau_\textrm{per}=15\tau$ and $\tau_\textrm{per}=63\tau$, respectively.  During the upwards motion, the liquid in the vicinity of the substrate is pushed upwards and mostly to the right. As the substrate velocity decreases, the strengths of the induced flow at the substrate decreases and, when the substrate reaches its maximal height, $\phi=\pi$, the average flow parallel to the substrate nearly vanishes at the substrate. When the substrate is moving down and starts to attract liquid the picture is different. We observe a pronounced average horizontal component of the velocity that points to the left. 
 
Additionally, we have calculated the average forces $\overline{F_{\textrm{up}}}$ and $\overline{F_\textrm{down}}$, that the substrate exerts on the liquid during the upward ($0 \leq \phi < \pi$) and downward ($\pi \leq \phi < 2\pi$) motion of the F-type substrate, respectively. These average forces are quantified by
\begin{equation}
 \overline{F_{\textrm{up}}} 
 = \frac{2}{\tau_\textrm{per}}  \Big\langle \int_0^{\tau_{\rm per}/2} \vec{F^{\textrm{s}}}(t) \cdot \vec{i} \, {\rm d}t \Big\rangle
\qquad \mbox{and} \qquad
 \overline{F_\textrm{down}} 
 = \frac{2}{\tau_\textrm{per}}  \Big\langle \int_{\tau_{\rm per}/2}^{\tau_{\rm per}}  \vec{F^\textrm{s}}(t) \cdot \vec{i} \, {\rm d}t \Big\rangle,
\end{equation}
where $\vec{F^\textrm{s}}$ is the force the substrate exerts on a liquid. The brackets $\big\langle \dots \big\rangle$ denote an ensemble average. The results are compiled in Tab.~\ref{tab:force_dir_motion}. The average force, $\overline{F_{\textrm{up}}} + \overline{F_\textrm{down}}$, is negative, resulting in a motion of the droplet to the left, and it decreases with $\tau_\textrm{per}$. 

\begin{table}[t]
	\centering
	\begin{tabular}{l|c|c|c}
	$\tau_\textrm{per}$ & $\overline{F_{\textrm{up}}}$, [$\sigma/\tau^2$] & $\overline{F_\textrm{down}}$, [$\sigma/\tau^2$] &  $\overline{F_{\textrm{up}}} + \overline{F_\textrm{down}}$, [$\sigma/\tau^2$]\\
	\hline
	$15\tau$ & $13.7 \pm 0.8$ & $-16.3 \pm 0.9$ & $-2.6$\\
	$63\tau$ & $5.3 \pm 0.5$ & $-5.6 \pm 0.6$ & $-0.3$\\
	\end{tabular}
	\caption{The average forces $\overline{F_{\textrm{up}}}$ and $\overline{F_\textrm{down}}$, acting on a droplet during the upward and downward motion of the F-type substrate at periods of $\tau_\textrm{per}=15\tau$ and $\tau_\textrm{per}=63\tau$. The average net force has a negative sign and therefore drives the droplet in the left direction.}
	\label{tab:force_dir_motion}
\end{table}

\begin{figure}[t]
    \centering
    \includegraphics[width=0.95\textwidth]{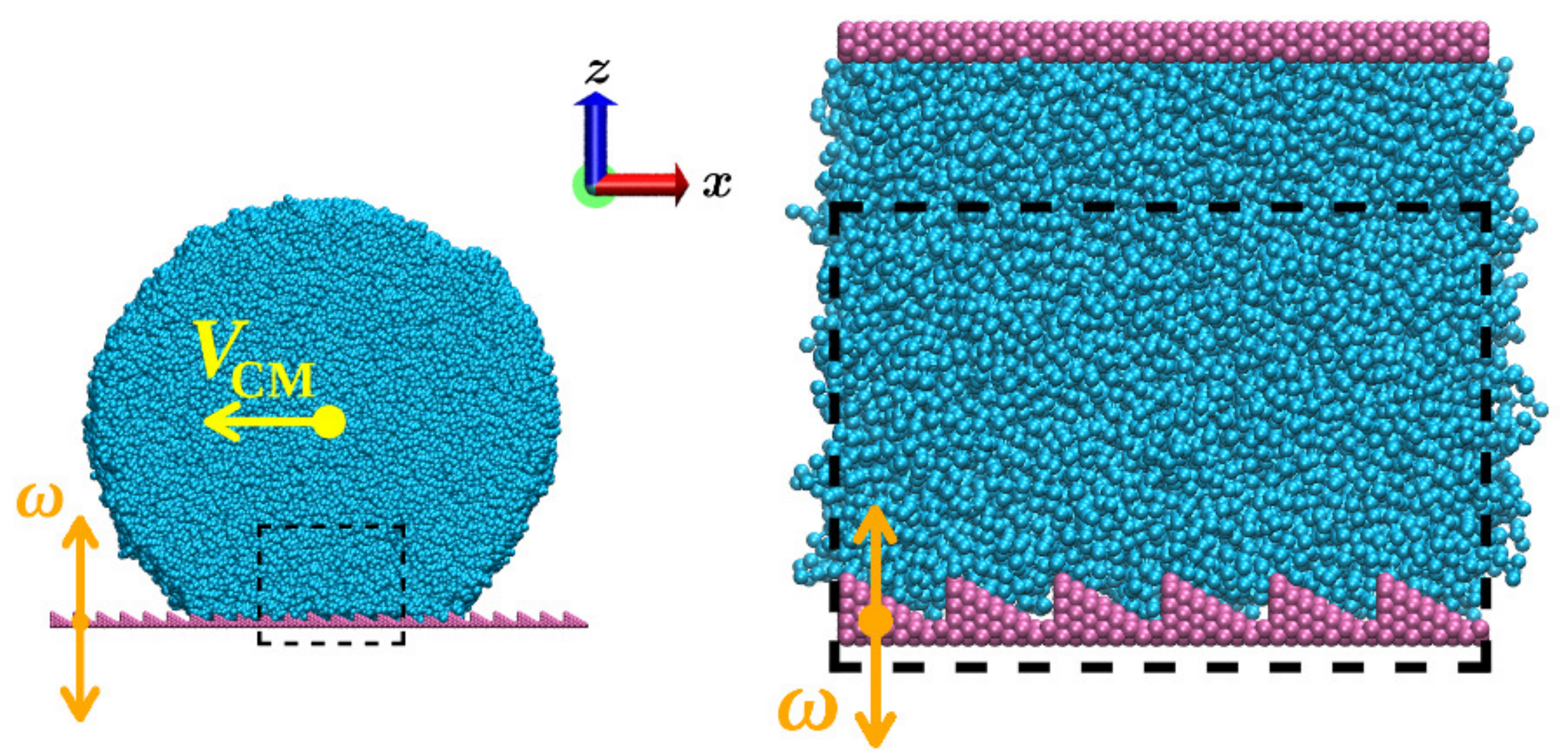}
    \caption{Side views of (a) a droplet on the ASVS of {the} F-type and (b) a flow of the liquid between a flat wall and the ASVS of {the} F-type. The dashed regions are equivalent, if the width of the channel is big enough to avoid an influence of the flat wall onto liquid's structure in the dashed box. The lack of advancing and receding CLs allows us to study the flows induced by the CA of the solid-liquid interface. If a flow in such a system exists, the CA with ASVS will drive the drop together with the CLs. If the flow is not developed, only the CLs may be the sources of the driving.}
    \label{fig:liq_under_osc}
\end{figure}

In order to investigate the driving of the CA without the influence of the CLs, we study a thin film (cf.~Fig.~\ref{fig:liq_under_osc} (right)). This set-up corresponds to a liquid volume close to the CA in the middle of a large drop as indicated by the dashed box in Fig.~\ref{fig:liq_under_osc} (left). In $z$ direction, the liquid is confined by a flat wall, whereas periodic boundary conditions are applied in $x$ and $y$ directions. The dashed regions of drop and film will be equivalent, if the packing of the liquid in the vicinity of the flat top wall does not influence the dashed box, and the density in the film and in the droplet coincide. Since the density of the vapor phase in pockets of the substrate is negligibly small and the coexistence pressure $p_\textrm{coex} \approx 0$~\cite{NT_MM_DT_UT_2012,NT_MM_slip_12}, we adjusted the density of the film to correspond to $p=0$. The vibrations of the asymmetric substrate in the channel are set up in the same way as for droplets.

\begin{figure}[t]
    \centering
    \subfloat[]{\label{fig:vel_prof_s3d0_1}\includegraphics[width=0.49\hsize]{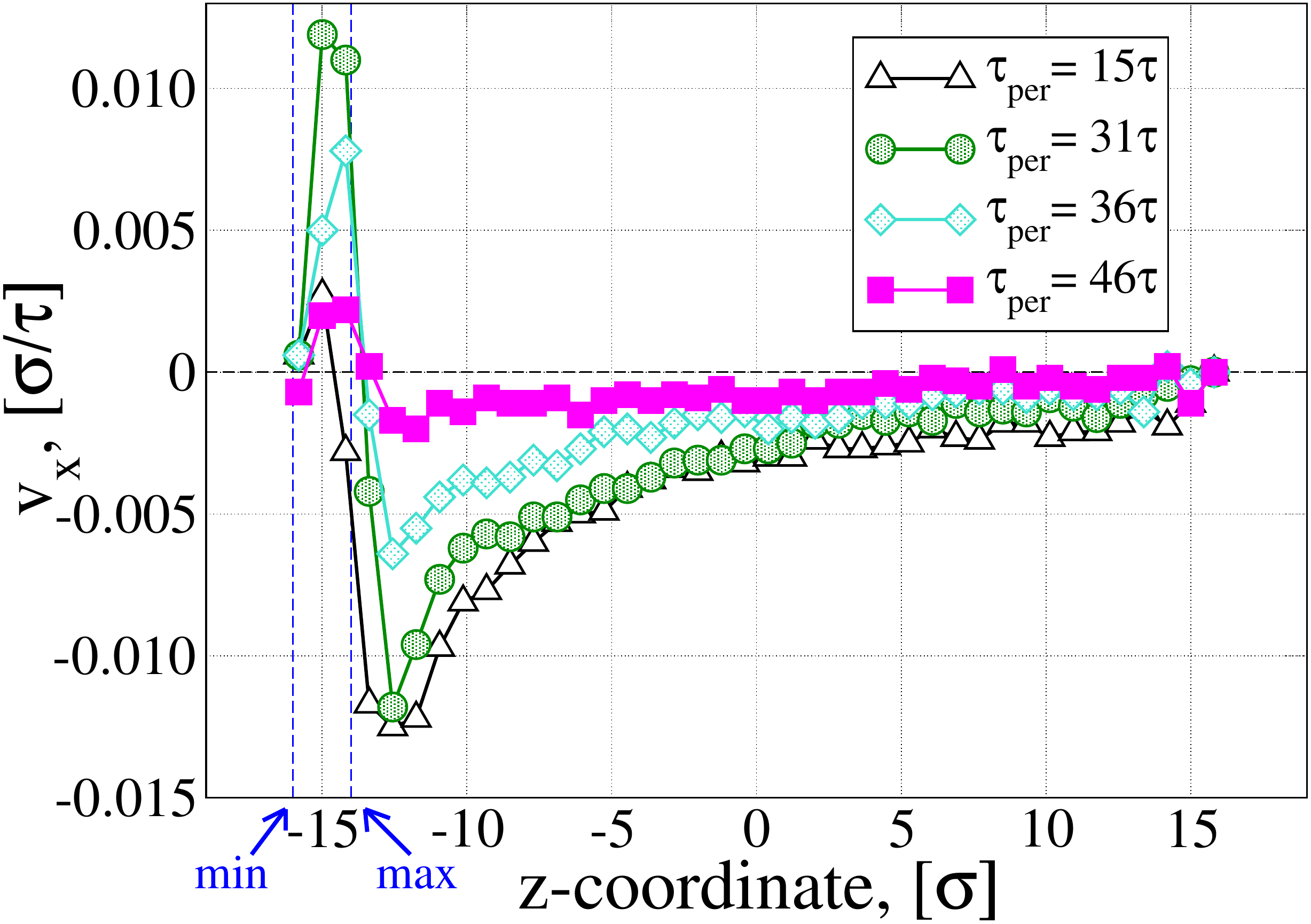}}\hspace{0.1cm}
    \subfloat[]{\label{fig:vel_prof_s7d0_1}\includegraphics[width=0.49\hsize]{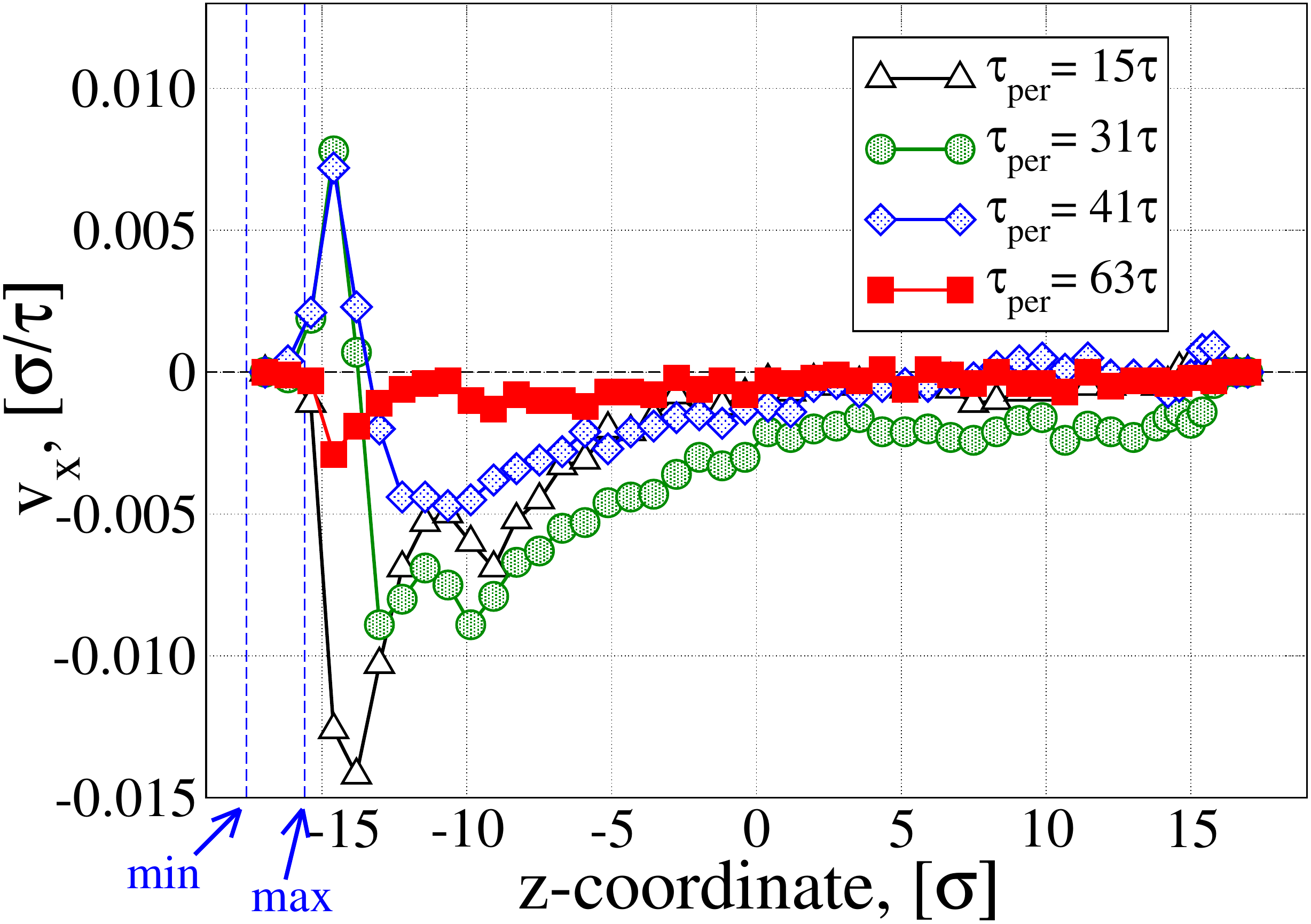}}
    \caption{Velocity profiles of the liquid confined between a flat and {the} F-type ASVS (a) and a flat and {the} R-type ASVS (b). The liquid is set in motion for periods of oscillations up to $\tau_\textrm{per} \leqslant 46\tau$ and $\tau_\textrm{per} \leqslant 63\tau$, for (a) and (b), respectively. The overshoots of the flow into a positive $x$ direction are dominated by the behavior inside the substrate cavities. {The minimum and maximum positions of the bottom-most layer of the substrate are indicated by arrows.}
    }
    \label{fig:vel_profiles}
\end{figure}

We measure the velocity profiles of the liquid as a function of the normal coordinate, $z$, at different periods $\tau_\textrm{per}$ for F- and R-type substrates. The simulation results in Fig.~\ref{fig:vel_profiles} clearly demonstrate that the substrate vibrations induce a net flow to the left (negative $x$-direction). This important observation indicates that the CA alone can drive the droplets at small $\tau_{\rm per}$. The driving is the strongest for small vibration periods, $\tau_{\rm per}$ and it lasts until $\tau_\textrm{per} \le 46\tau$ and $\tau_\textrm{per} \le 63\tau$ for F- and R-type substrates, respectively. 

\section{Mechanisms of droplet motion}
\mylab{sec:agit:mech}

\subsection{Steady-state center-of-mass velocity and power balance}
Both the stick-slip motion of the CLs and, at short vibration periods, the {motion of the} CA contribute to drive the droplets to the left. In this section we study the velocity of the droplet's center of mass in steady state as a function of droplet size and vibration period, $\tau_\textrm{per}$, and rationalize our observations by studying the different dissipation mechanisms.

\begin{figure}[t]
	\includegraphics[width=0.9\textwidth]{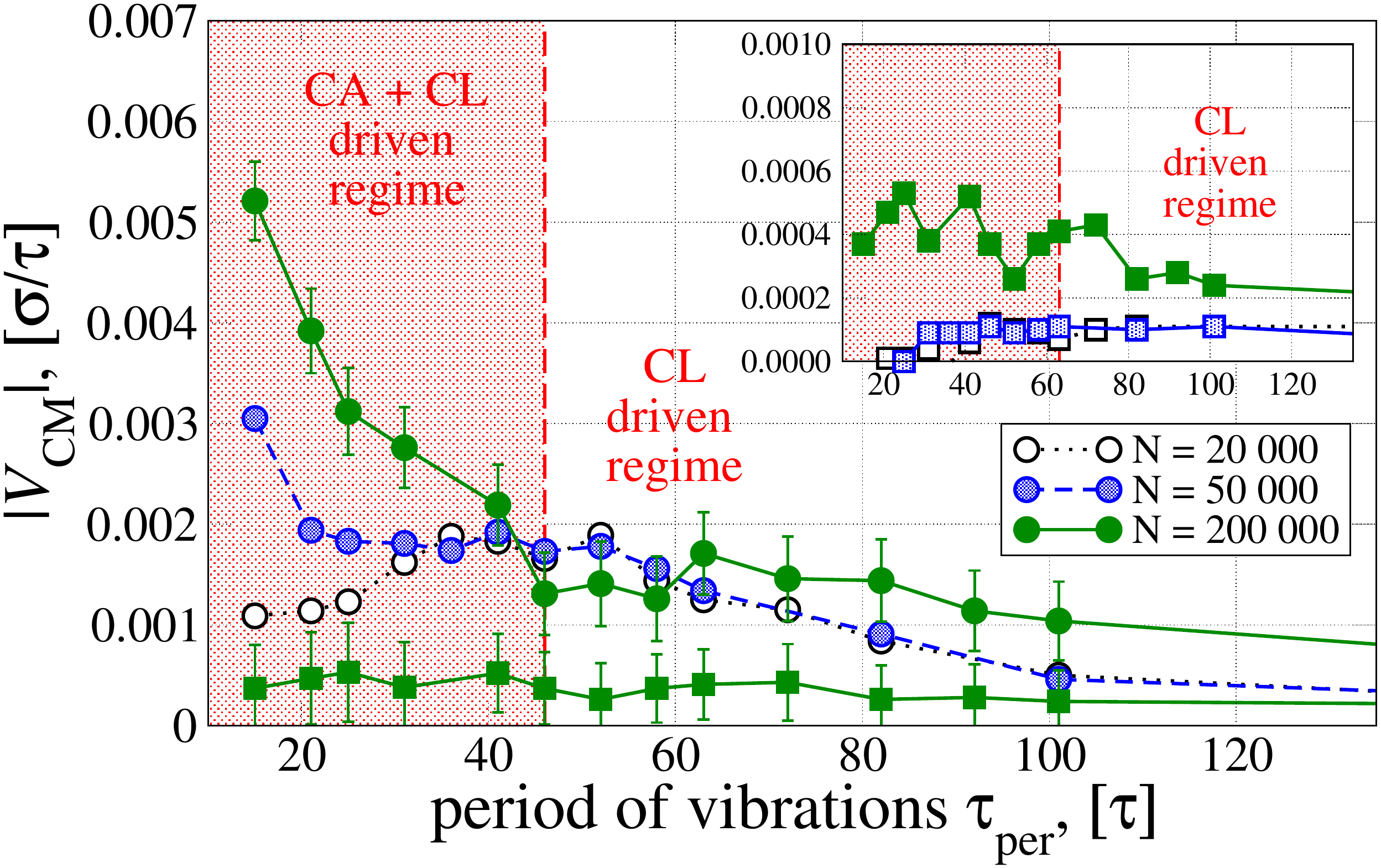}
	\caption{Velocity of the centre-of-mass, $V_\textrm{CM}$, for drops of various size as a function of the period of F-type (main panel) and R-type (inset) substrate vibrations. For comparison, $V_\textrm{CM}$ of the biggest droplet at R-type substrate is also plotted on the main panel (filled squares). For sake of clarity we plot the error bars only for the largest droplets. The shaded regions mark the regime of joined activity of the CA and CLs. In the main panel, two mechanisms of driving clearly visualize the difference in the velocity of the CM: if the CLs drive the droplet, the size of the droplet does not influence its CM velocity significantly; if the CA, however, also an active mechanism, the discrepancy in the velocity of the CM with droplet size is present. At R-type substrate (the inset) the latter effect is affected by a very small CA of drops of $20\,000$ and $50\,000$ beads. Both drops behave similarly and their velocities are hardly different from zero. For the details see the main text.}
    \label{fig:velCM}
\end{figure}

The main panel of Fig.~\ref{fig:velCM} presents the extensive simulation data for the F-type substrate, whereas the data for the R-type corrugation are depicted in the inset. For sake of comparison we additionally include the much smaller velocity of the largest droplet on the R-type substrate (filled squares) in the main panel. The error bars are estimated from the variation of $15-20$ independent simulation runs.

First, we focus on the F-type substrate. Fig.~\ref{fig:velCM} demonstrates that at large vibration periods, where only the CLs drive the droplet (non-shaded region), its center-of-mass velocity, $V_\textrm{CM}$, is small and decreases with increasing $\tau_{\rm per}$. Within the statistical accuracy of the simulation data, the size of the droplet has no influence on its velocity $V_\textrm{CM}$. 

At short periods of vibrations the CA and CLs are simultaneously active in driving the droplet (shaded region). In this regime,  $V_\textrm{CM}$ increases with decreasing $\tau_{\rm per}$ and it increases with the size of the droplet. Therefore, vibrations of the asymmetrically structured substrate at short periods provide a mechanism for sorting droplets according to their size, i.e., the larger the droplet is, the farther it moves on the vibrating substrate in a given time.

The absence of pronounced vapor pockets on the F-type substrate results in a larger contact between the liquid and the solid than on the R-type substrate. Thus the CA driving is stronger on the F-type substrate than on the R-type one and we observe that $V_\textrm{CM}$ is larger (filled circles {\it vs.} filled squares in the main panel of Fig.~\ref{fig:velCM}).

The center-of-mass velocity in steady state is dictated by a balance between the power that the ASVS imparts into the directed motion and its dissipation during the motion~\cite{deGennes_1985,Brochard_1989, JS_MM_2008, DB_JE_JI_09}. First, we quantify the power input provided by the substrate vibration. 

Following {\it de Gennes}~\cite{deGennes_1985}, we formulate the balance of the energy input and its dissipation per unit time. The input power provided by the substrate vibrations, $P_\mathrm{in}$, will be dissipated by several channels. For a liquid wedge, originally~\cite{deGennes_1985}, viscous dissipation, $T\Sigma_\textrm{V}$, of the fluid flow inside the droplet, dissipation at the three-phase CLs, $T\Sigma_\textrm{CL}$, and in the precursor film, $T\Sigma_{\rm prec}$ have been considered~\cite{LM_YP_1999}. Since our solid is hydrophobic, the thermodynamic conditions are remote from the strongly first-order wetting transition, and no precursor film is observed, i.e., $T\Sigma_{\rm prec} = 0$. Since the fluid may slip past the substrate, however, frictional dissipation at the CA, $T\Sigma_\textrm{CA}$, must be considered~\cite{JS_MM_2008}. 
Moreover, the vibrating solid induces periodic density compressions leading to sound waves, whose energy will be dissipated due to their damped propagation in the liquid. We denote this additional dissipation phenomena by $T\Sigma_\textrm{SW}$. We note that estimating the strength of these dissipation mechanisms relies on a macroscopic description (e.g., involving the hydrodynamic velocity profile without thermal fluctuations to compute the viscous dissipation $T\Sigma_\textrm{V}$ {and dissipation due to damping of the sound waves, $T\Sigma_\textrm{SW}$}).  

The power, $ P_\mathrm{in}$, imparted by the vibrating substrate onto the the system, however, is computed microscopically from the instantaneous fluctuating force between the solid and the liquid. {\it A priori} it is not obvious that all this input power will contributes to the driving of the droplet -- a portion of this microscopic input power may be directly dissipated into heat and removed by the thermostat. Therefore we expect an additional dissipation term, $T\Sigma_\textrm{F}$, accounting for this direct conversion of vibration into microscopic thermal fluctuations. We expect this term to be particularly relevant at short vibration periods, where $\tau_{\rm per}$ starts to become comparable to the characteristic time scales of the liquid, $\tau$. 

Thus, the balance between input power and dissipation during the motion of the droplet is: 

\begin{equation}
 P_\mathrm{in} = \overbrace{T\Sigma_\textrm{V}}^\text{3D} +  \underbrace{T\Sigma_\textrm{SW} + T\Sigma_\textrm{CA} + T\Sigma_\textrm{F}}_\text{2D} + \overbrace{T\Sigma_\textrm{CL}}^\text{1D}.
 \label{eq:drop_dissip}
\end{equation}

It is important to note that all dissipation terms of Eq.~(\ref{eq:drop_dissip}) are characterized by their spatial dimension:  {Viscous dissipation}, $T\Sigma_\textrm{V}$,  occur in the 3-dimensional volume of the droplet that scales like $R^2 L_y$ with $R$ denoting the droplet radius. {Sound-wave, frictional, and fluctuation dissipation},  $T\Sigma_\textrm{SW}$, $T\Sigma_\textrm{CA}$ and $T\Sigma_\textrm{F}$, take place at the vicinity of the solid-liquid interface. They are associated with a 2-dimensional area that scales like $\sim R L_y$. {CL dissipation}, $T\Sigma_\textrm{CL}$, involves the 1-dimensional contact lines of length $\sim 2L_{y}$.

In the following subsections we study the {input power} and individual dissipation mechanisms of Eq.~(\ref{eq:drop_dissip}) in turn.

\subsection{Input power}
\mylab{sec:dissip:input}

\begin{figure}[t]
	\subfloat[]{\label{fig:sub_input_s3d0}\includegraphics[width=0.8\hsize]{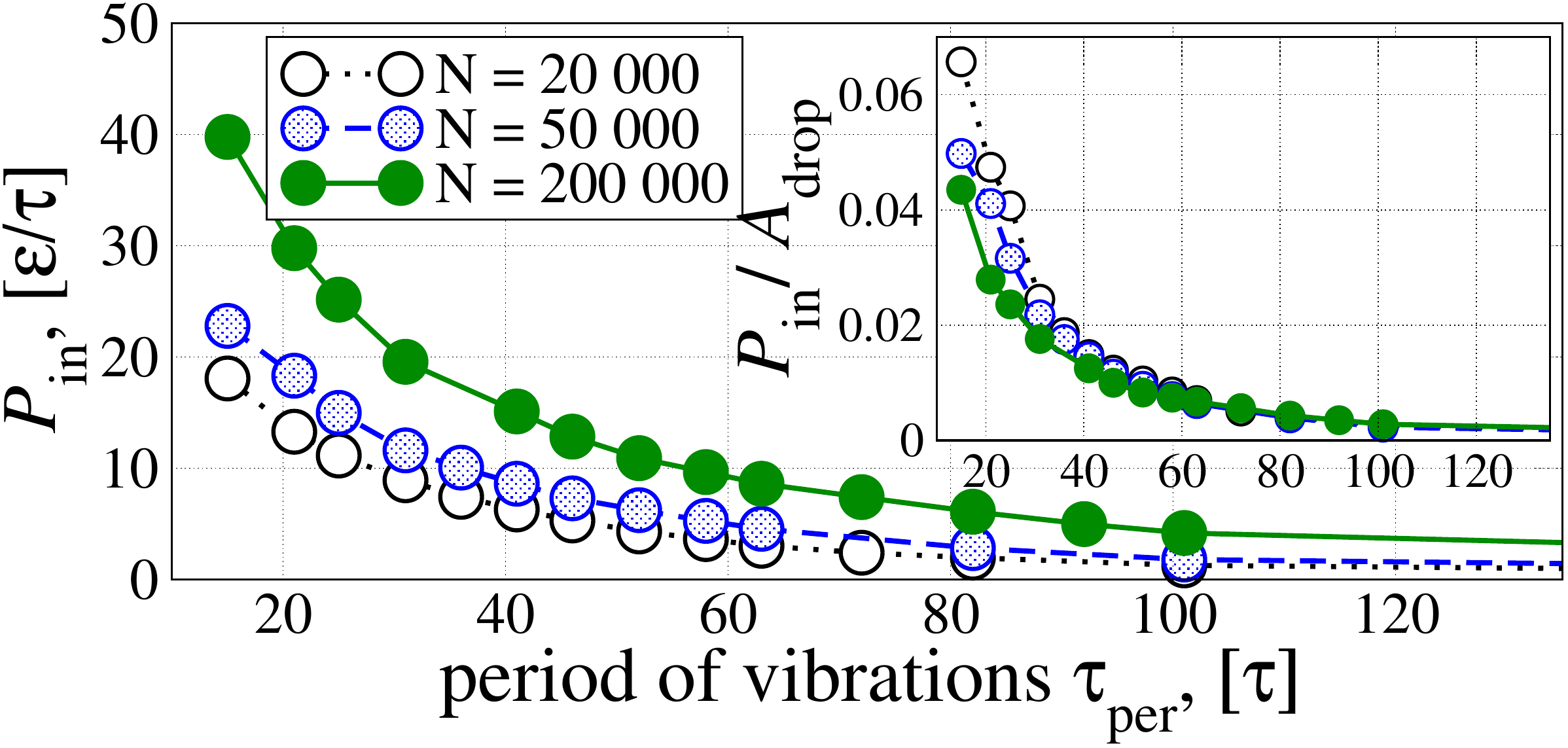}}
	
	\subfloat[]{\label{fig:sub_input_s7d0}\includegraphics[width=0.8\hsize]{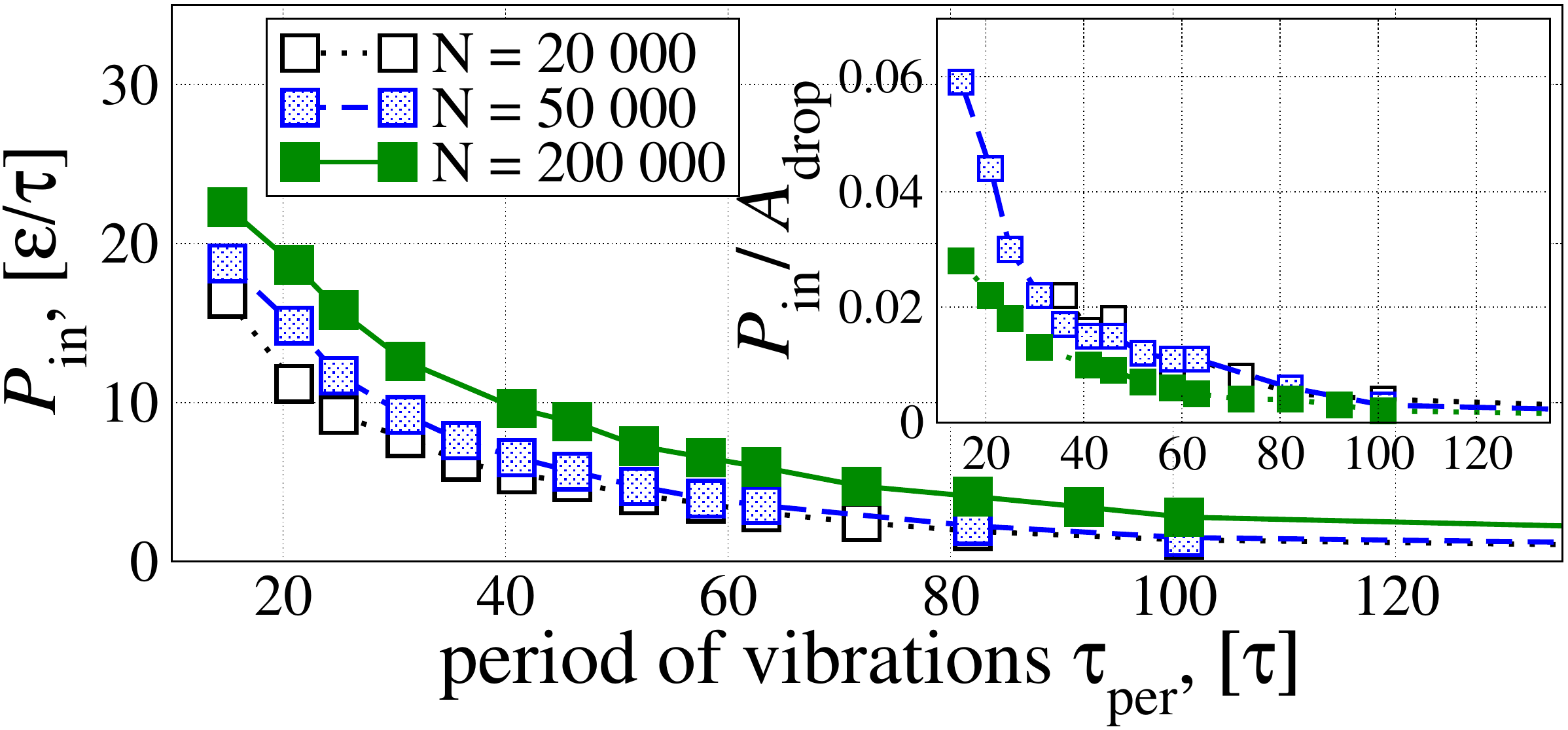}}
	\caption{The input power $P_\mathrm{in}$ of vibration of F-type (a) and R-type (b) substrates as a function of the period $\tau_\textrm{per}$ of vibration. Different colors (and symbols shadings) correspond to various drop sizes $N$. The common trend is the decreasing of the power with the period. The inset shows the scaling of the input power with the contact area of the droplet. At short periods the specific power for smaller droplets turns out to be somewhat greater than for bigger drops. At greater periods the dependencies tend to coincide (see main text for details). }
    \label{fig:power_in_per}
\end{figure}

The power input can be expressed in terms of the instantaneous deterministic force 
\footnote{The dissipative and random forces of the thermostat are not considered in computing the input power. The thermostating scheme merely impose the temperature $T$ provided that the friction constant of the DPD-thermostat is sufficiently large to avoid the overheating of the liquid in contact with the vibrating substrate. Our choice of $\Gamma = 0.5/\tau$~\cite{PE_PW_1995,FL_JS_CP_MM_2011} fulfills this criterion well even at the shortest period of substrate vibrations of $\tau_\textrm{per}=15\tau$.}
exerted by the ASVS onto a droplet, $\vec{F^\textrm{s}}(t)$, and the velocity of the substrate:
\begin{equation}
	P_\mathrm{in}= \frac{1}{\tau_\textrm{per}} \Bigg\langle \int_0^{\tau_\textrm{per}} v_z^{\rm{s}}(t) \, \vec{F^\textrm{s}}(t) 	\cdot \vec{n}_{z} \, {\rm d}t \Bigg\rangle.
    \label{eq:pow_sub_vib}
\end{equation}
The velocity of the substrate, $v_z^{\rm{s}}(t) \sim \frac{1}{\tau_{\rm per}}$, decreases with the period of oscillations, which is confirmed in Fig.~\ref{fig:power_in_per} for both types of substrates. Moreover, $P_{\rm in}$ is proportional to the CA because the force in Eq.~(\ref{eq:pow_sub_vib}) scales like the number of liquid-solid interactions. The CA increases very slightly with the period giving rise to small deviations from the $1/\tau_{\rm per}$-dependency. 

In the insets of Fig.~\ref{fig:power_in_per}, we present the input power per projected contact area $A_{\rm drop}$. This ratio, $P_{\rm in}/A_{\rm drop}$ is almost independent from the droplet radius $R$ at large $\tau_{\rm per}$ but slightly decreases with $R$ at smaller periods. This observation might be partially rationalized by an additional contribution in the narrow area in the vicinity of the fluctuating contact line, which becomes relatively more important for small droplets. We propose a relation of the form
\begin{equation}
P_\textrm{in} = p_\textrm{spec} (A_\textrm{drop} + \Delta x L_y),
\label{eq:p_in}
\end{equation}
where $\Delta x$ is the effective width of the fluctuating CL region. Since the CL fluctuations are more prominent at small $\tper$ than at larger periods, $\Delta x$ decreases with $\tper$ in agreement with the size-dependence of $P_{\rm in}/A_{\rm drop}$ presented in the insets of Fig.~\ref{fig:power_in_per}. Comparing F-type and R-type substrates we note that the size-dependence of $P_{\rm in}/A_{\rm drop}$ at small $\tper$ is larger on the R-type substrate. This finding is in accord with macroscopic considerations suggesting that the magnitude of CL fluctuations is related to the corrugation. This implies large $\Delta x$ for the R-type substrate.

Additionally we note that $P_{\rm in}/A_{\rm drop}$ is larger for the F-type substrate than for the R-type substrate. This observation is rooted in the microscopic corrugation of the substrate. Whereas macroscopic considerations suggest that the actual area of contact between liquid and solid is independent from the length scale of the corrugation, we observe that there are no significant vapor pockets on the F-type substrate but that these cavities exist on the R-type substrate. Thus the actual contact area and thereby also the number of liquid-solid interactions and $\vec{F^\textrm{s}}$ of Eq.~(\ref{eq:pow_sub_vib}) is larger on the finely corrugated F-type substrate than on the R-type one.

\subsection{Viscous dissipation}
\mylab{sec:dissip:Vic}

The viscous dissipation in the droplet with hydrodynamic velocity field, $\vec{u}$, is defined as~\cite{LL_EL_87}
 \begin{equation}
 T\Sigma_\textrm{V} = \frac{1}{2} \eta \int_{V} \bigg(\frac{\partial v_k}{\partial x_i} + \frac{\partial v_i}{\partial x_k} - \frac{2}{3}\delta_{ik}\frac{\partial v_l}{\partial x_l}\bigg)^2 \, dV. 
 \label{eq:visc_diss}
 \end{equation}
where $V$ is the volume, over which the viscous dissipation occur and the Einstein summation convention is implied. The shear viscosity of the liquid is $\eta=5.32 \pm 0.09 \sqrt{m\epsilon}/\sigma^2$ is defined from the shear stress autocorrelation function~\cite{FL_JS_CP_MM_2011}.

Eq.~(\ref{eq:visc_diss}) is valid for a liquid bulk and is defined at a macroscopic level (i.e., using a velocity field without fluctuations). In order to evaluate the integral in Eq.~(\ref{eq:visc_diss}), we use the ensemble averages of the velocity field in a droplet at various phases during the substrate vibration. As a compromise between available data storage, speed of analysis and its accuracy we use the same four phases of harmonic oscillations, as before in Sec.~\ref{sec:agit:response}. The velocity fields are then discretized to a regular grid and Eq.~(\ref{eq:visc_diss}) is evaluated by finite differences. The grid size is chosen as to minimize statistical errors and effects of thermal fluctuations. Grid spacings vary between {$2 \sigma$} and $8\sigma$ yielding similar results. We note that vibrations give rise to small density variations and therefore $\partial v_l / \partial x_l \neq 0$.

\begin{figure}[t]
	\subfloat[]{\label{fig:visc_dis_s3d0}\includegraphics[width=0.8\hsize]{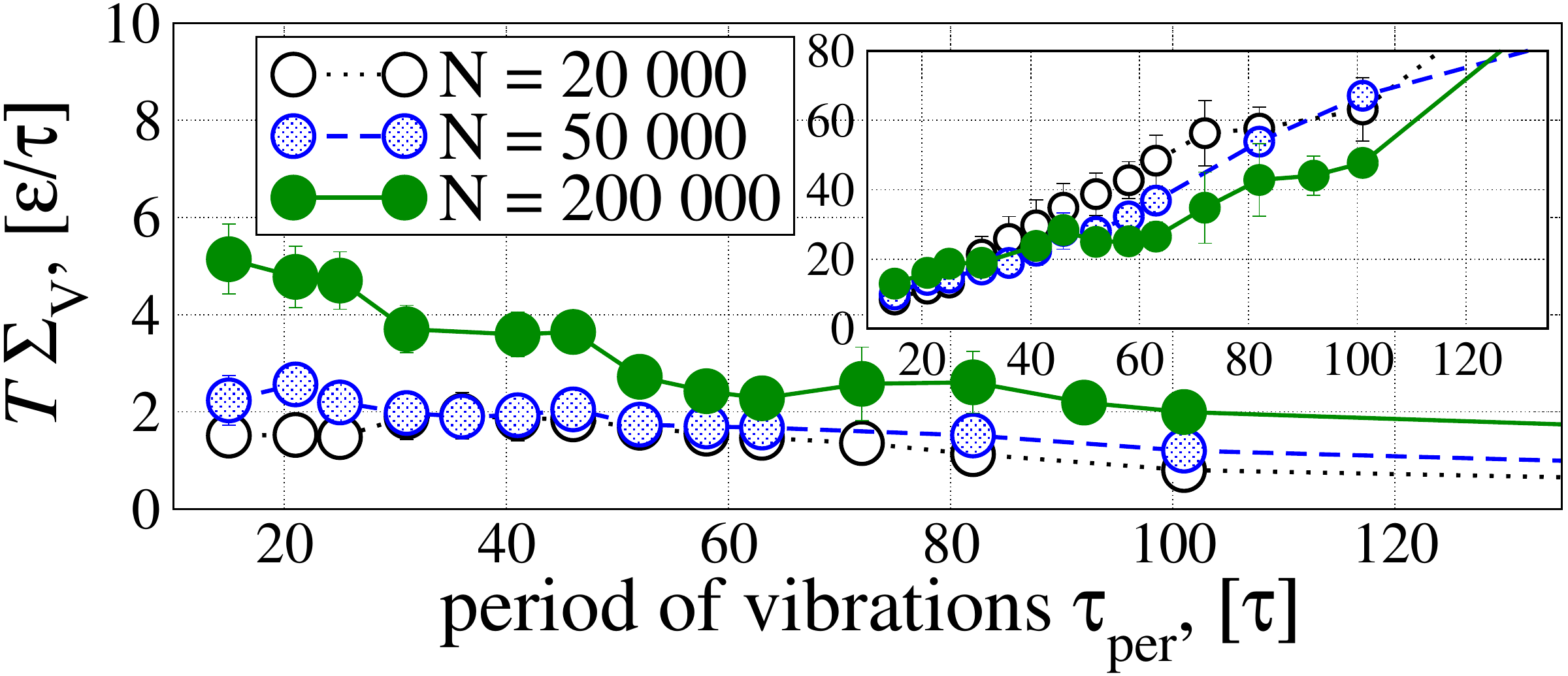}}
	
	\subfloat[]{\label{fig:visc_dis_s7d0}\includegraphics[width=0.8\hsize]{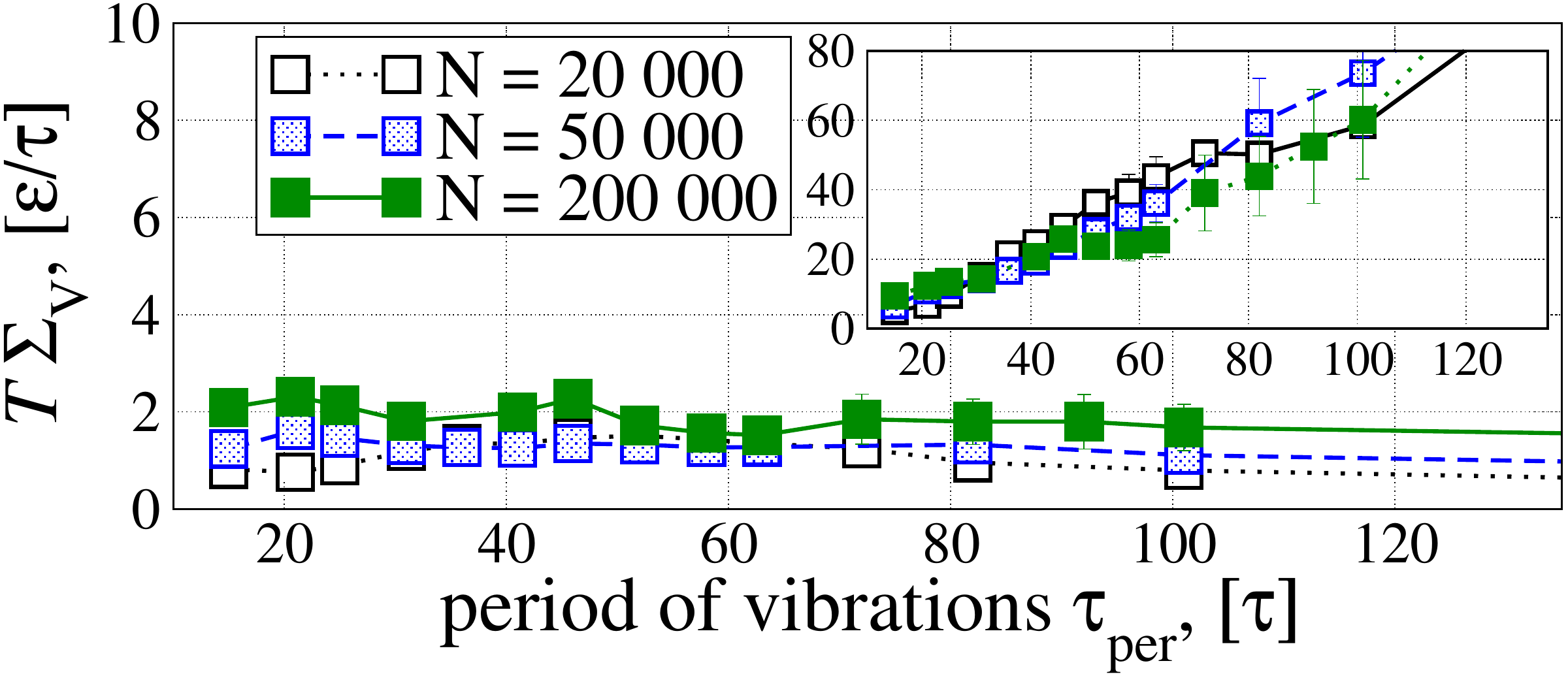}}
	\caption{The viscous dissipation $T\Sigma_\textrm{V}$ of drops on ASVS of F- and R-type, (a) and (b), correspondingly, as functions of the period $\tau_\textrm{per}$. Different colors (and symbols shadings) correspond to various drop sizes $N$. The dissipation rate does slightly decrease with period, as the strength of the flow inside the droplet is descending. The drops of the same size looses more energy on a finely corrugated substrate (F-type) than at the substrate of R-type. For explanation see the main text. The insets represent the relative strength (in percent) of viscous dissipation in comparison to the power input.}
    \label{fig:visc_diss}
\end{figure}

Fig.~(\ref{fig:visc_diss}) presents the results of Eq.~(\ref{eq:visc_diss}) on both substrate types. The viscous dissipation $T\Sigma_\textrm{V}$ is larger for larger droplets at given $\tper$ because (i) the volume in which the liquid flow dissipates is larger and (ii), at small $\tper$, the larger droplets additionally move faster resulting in larger shear forces inside the liquid. This is also one of the reasons why the viscous dissipation is larger for the faster moving droplets on the F-type substrate than for the R-type substrate. Additionally, the flow pattern inside the droplet differs between F-type and R-type substrates.

\begin{figure}[t]
	\begin{tabular}{M|M|M}
	Substrate & $\tau_\textrm{per}=15\tau$ & $\tau_\textrm{per}=63\tau$ \\
	\hline
	F-type & \includegraphics[width=0.4\hsize]{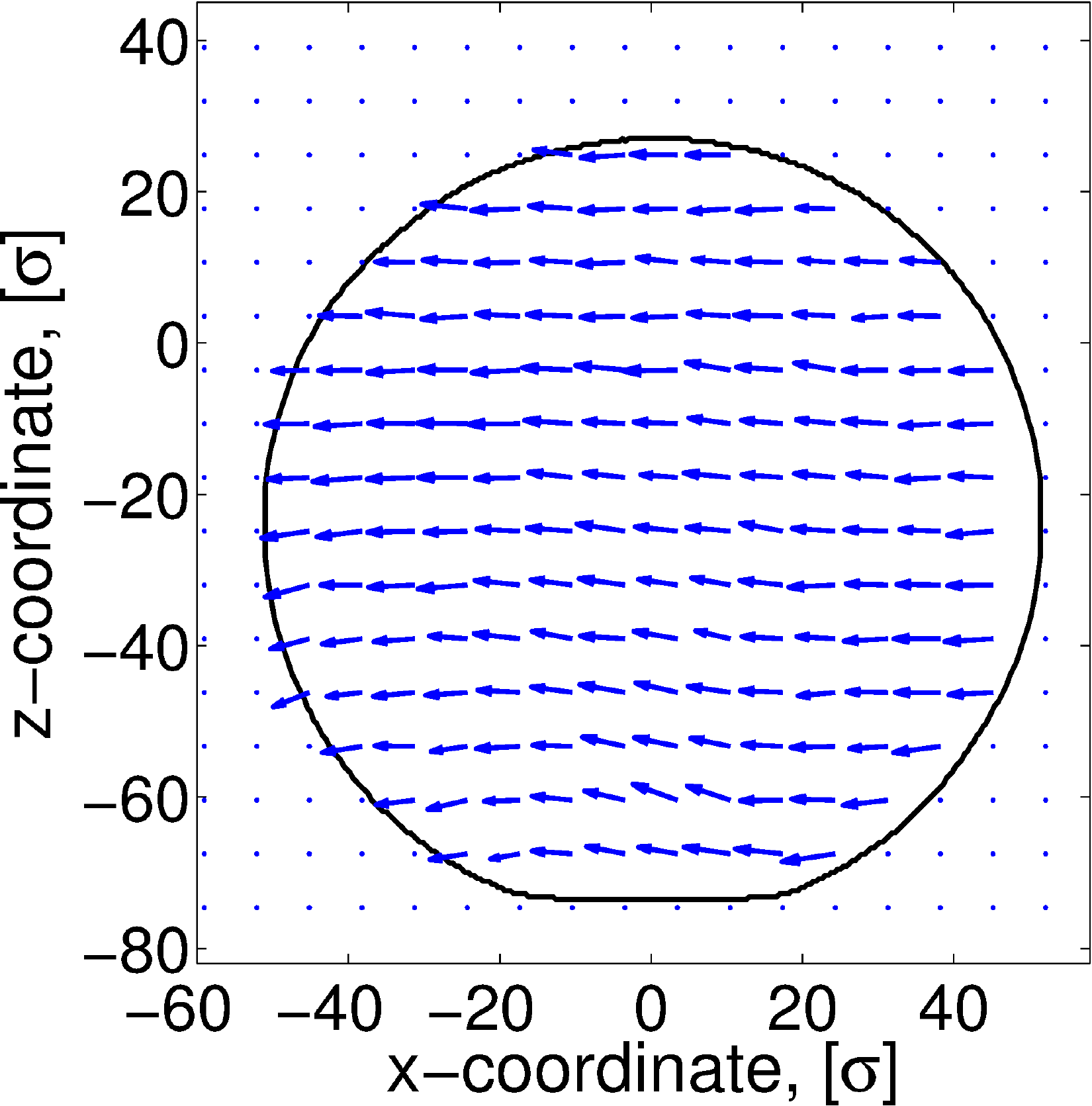} & \includegraphics[width=0.4\hsize]{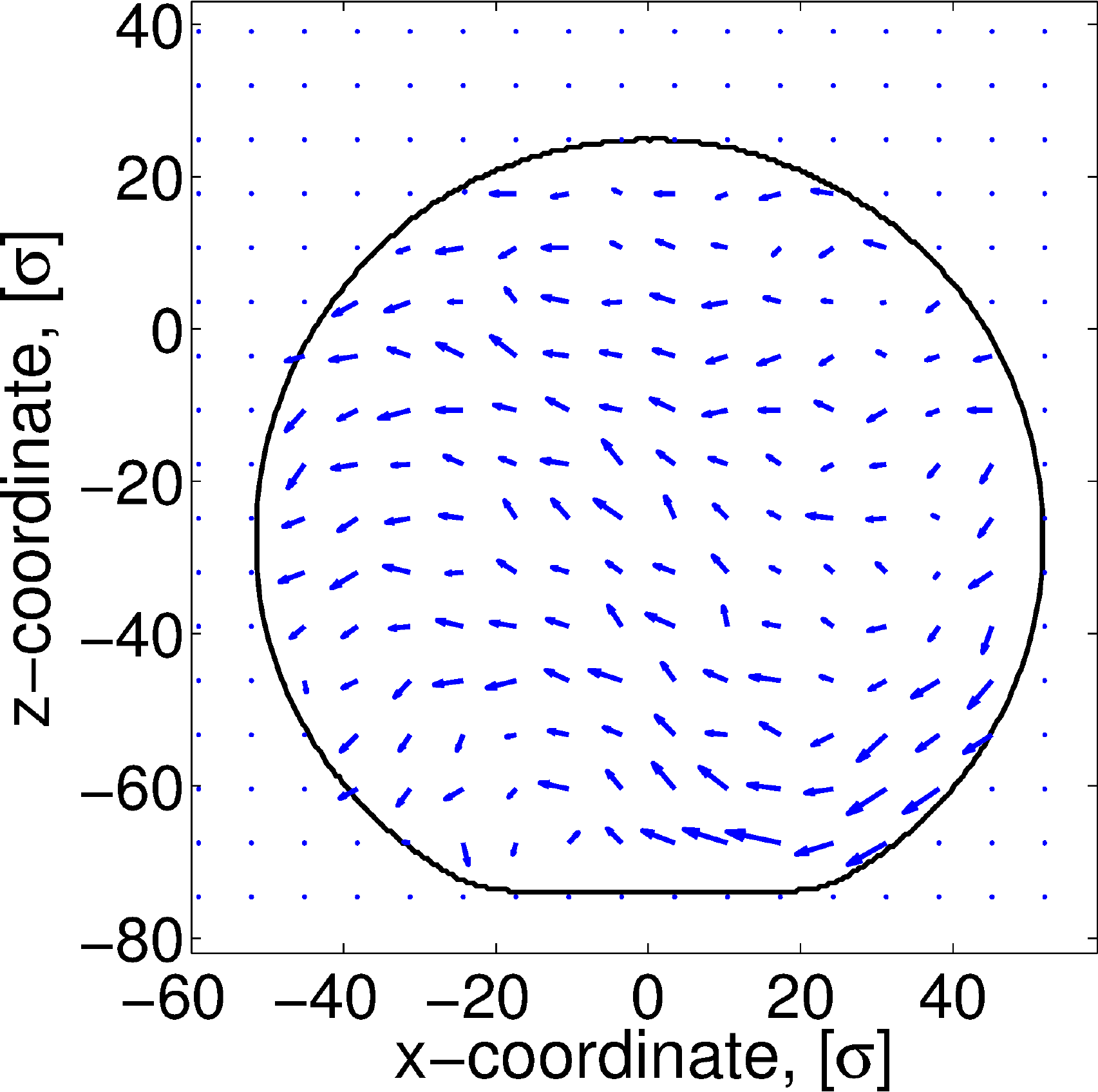} \\
	\hline
	R-type & \includegraphics[width=0.4\hsize]{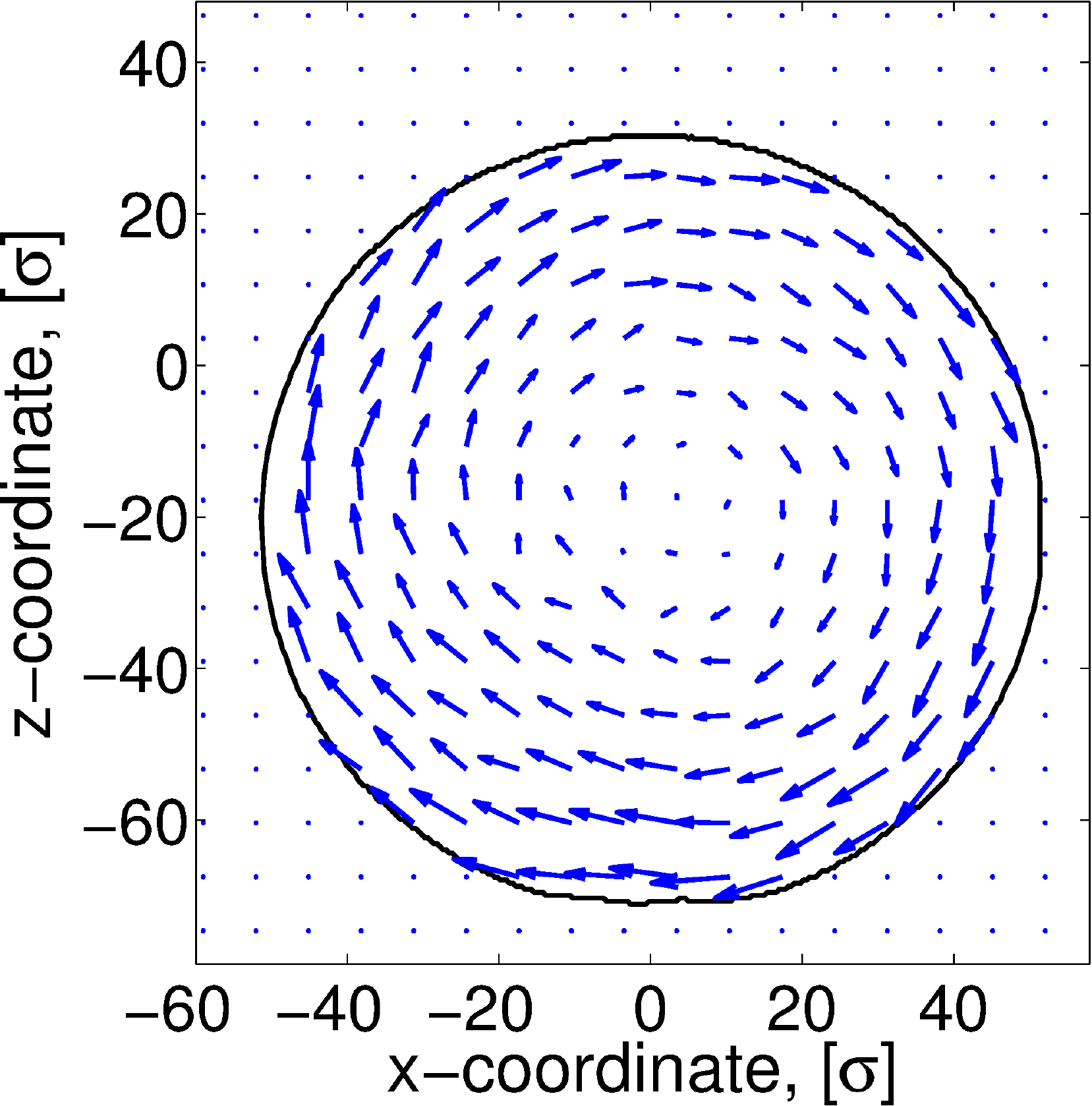} & \includegraphics[width=0.4\hsize]{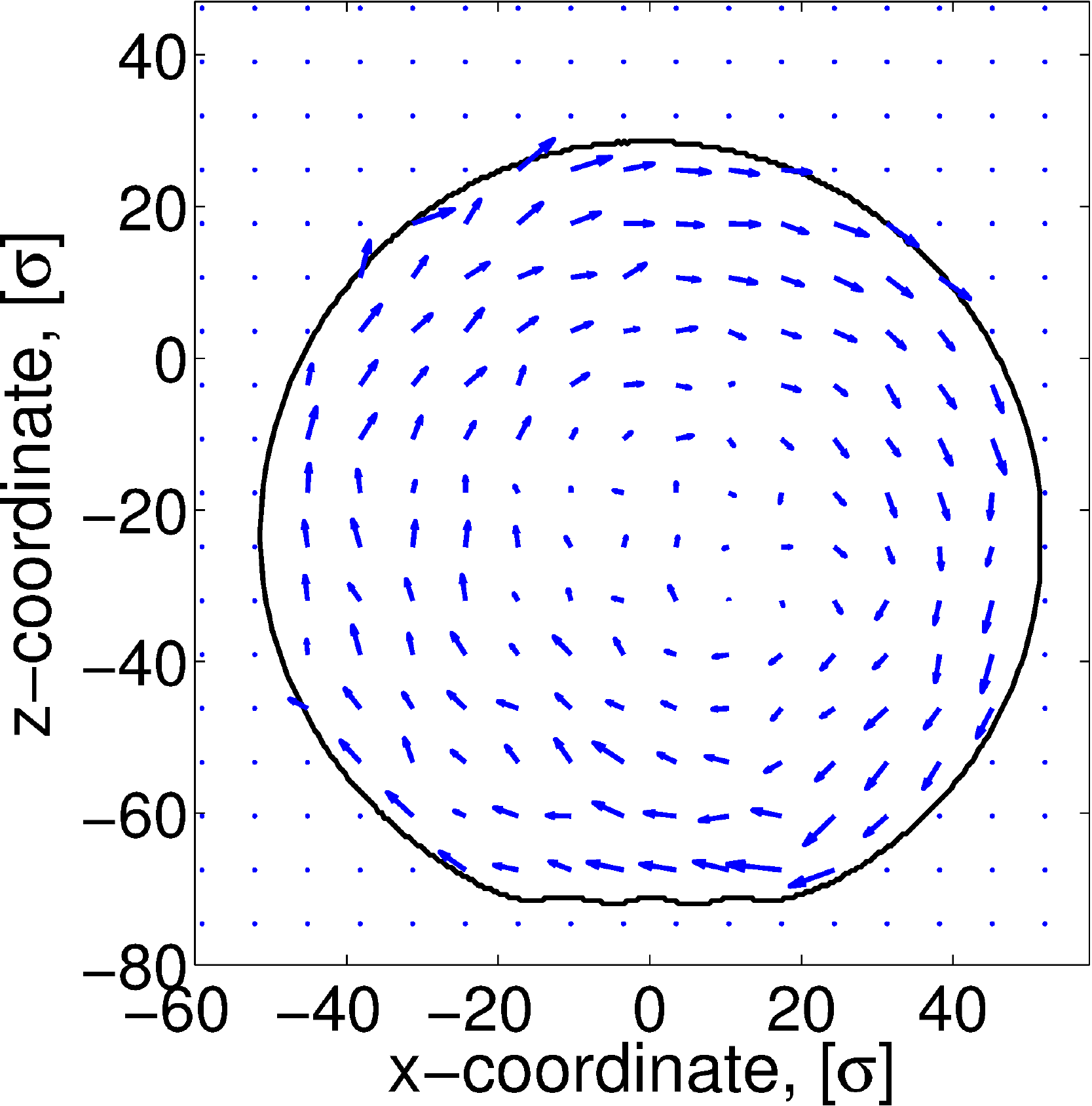} \\
	\end{tabular}
	\vspace*{0.3cm}
	\caption{Velocity fields of moving droplets of $N = 200\, 000$ beads at the F-type substrate (top row) and the R-type substrate (bottom row). The left column stands for vibration period of $\tau_\textrm{per}=15\tau$, the right one for the period of $\tau_\textrm{per}=63\tau$. All velocity field are shown in a LAB system (from the point of view of an observer). The difference in the flow patterns may be explained by the impact of the substrate onto the liquid by vibration (see the main text for details).
	}
	\label{fig:vel_fields_FRsub}
\end{figure}

Averaging the velocity fields over $40 \tper$, we  obtain an averaged macroscopic velocity field in Fig.~\ref{fig:vel_fields_FRsub} for both substrate types and two values of $\tper=15\tau$ and $63\tau$. The former corresponds to the regime where both CA and CLs drive the droplet, while in the later regime only the stick-slip motion of the CLs is active. The data refer to the largest droplet size, $N = 200\, 000$, but the behavior of the smaller droplets is qualitatively similar. 

The scale of the substrate corrugation has a pronounced influence on the time-average flow field. On the roughly corrugated substrate, the droplets roll. As we increase $\tper$ the droplets move slower, giving rise to an overall smaller scale of the velocity, but the rolling character of the fluid flow persists. We also note that the clockwise direction of rotation is opposite to the one that the rigid cylinder would have rolling to the left because the driving force is localized at the substrate. 

On the finely corrugated substrate, however, the fluid is mainly sliding at small $\tper$, whereas at larger $\tper$ a more complex average flow pattern is evident. The difference between the flow patterns on the structured substrates under investigation may be explained by the impact of the substrate onto the liquid. While there are hardly any vapor pockets and the edges of the grooves are separated by $4.81\sigma$ for the F-type substrate, this distance increases to $9.62\sigma$ for the R-type one. Therefore, in the former case the impact of a substrate vibrations is more uniformly distributed over the contact area, while in the latter case there is rather a collection of a separated impact points (the positions of the edges) giving rise to an additional rotation. 

The insets of Fig.~\ref{fig:visc_diss} present the relative contribution of the viscous dissipation $T\Sigma_\textrm{V}$ with respect to the power input, i.e., the ratio $T\Sigma_\textrm{V} / P_\textrm{in}$,  as a function of the period of vibrations. At small $\tper$ viscous dissipation is only a small fraction of the total energy input but it becomes increasingly important at larger $\tper$. 

\subsection{Dissipation by sound waves}
\mylab{sec:dissip:sound}  

\begin{figure}[t]
    \centering
    \subfloat[]{\label{fig:sound_s3d0_n19712_e04_15_ph2}\includegraphics[width=0.46\hsize]{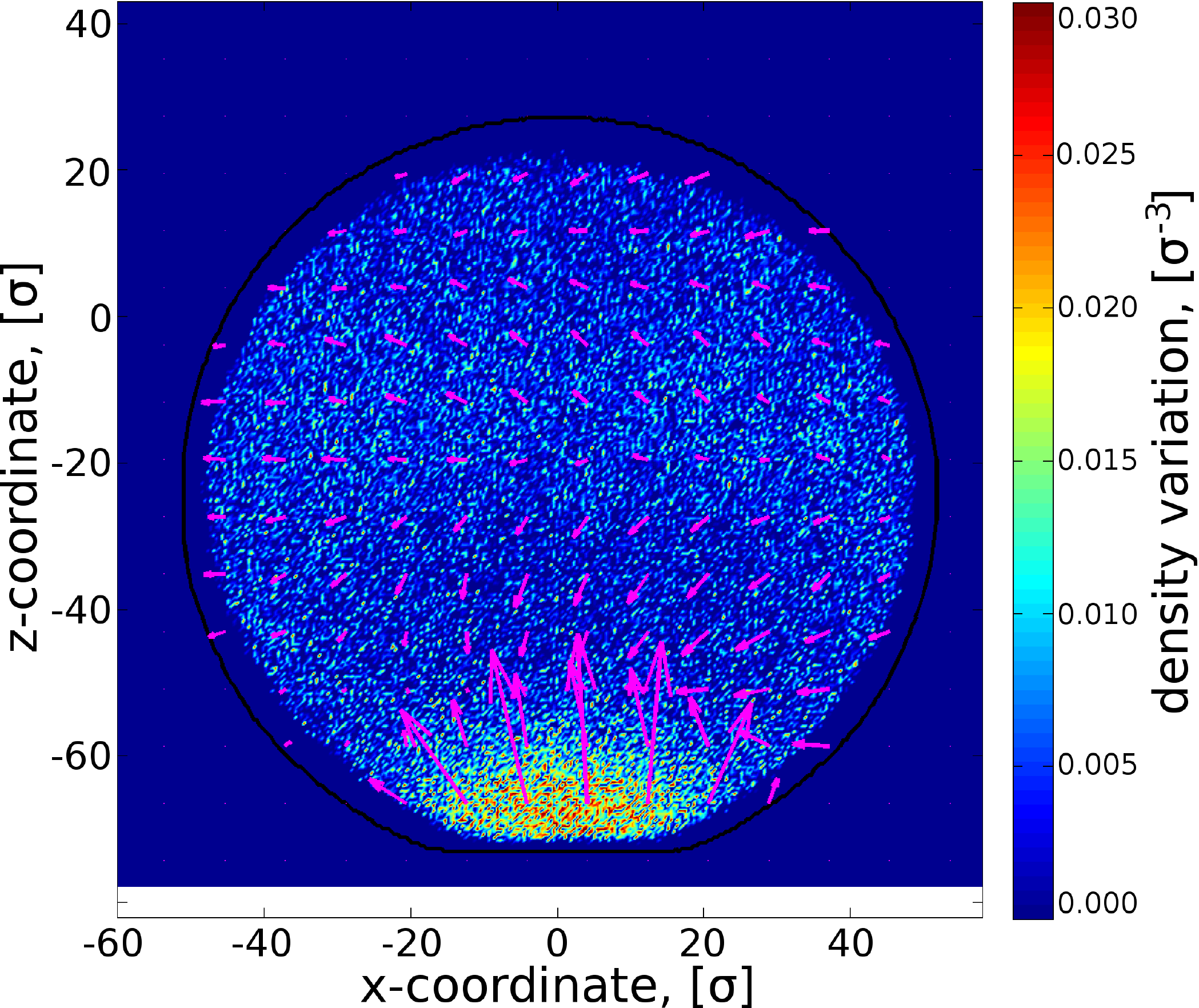}}\hspace{0.1cm}
    \subfloat[]{\label{fig:sound_s3d0_n19712_e04_15_ph3}\includegraphics[width=0.46\hsize]{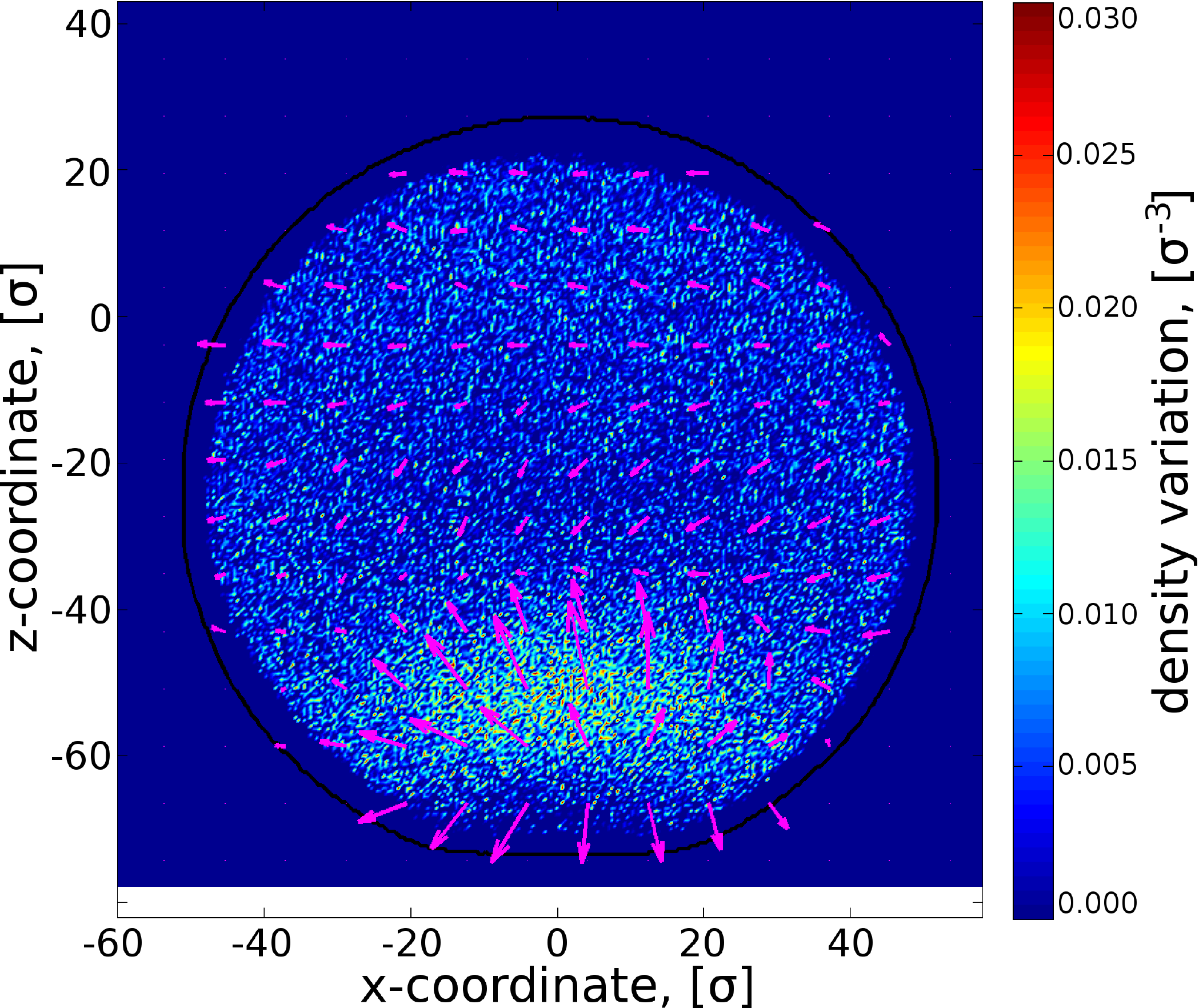}}\hspace{0.1cm}
    
    \subfloat[]{\label{fig:sound_s3d0_n19712_e04_15_ph0}\includegraphics[width=0.46\hsize]{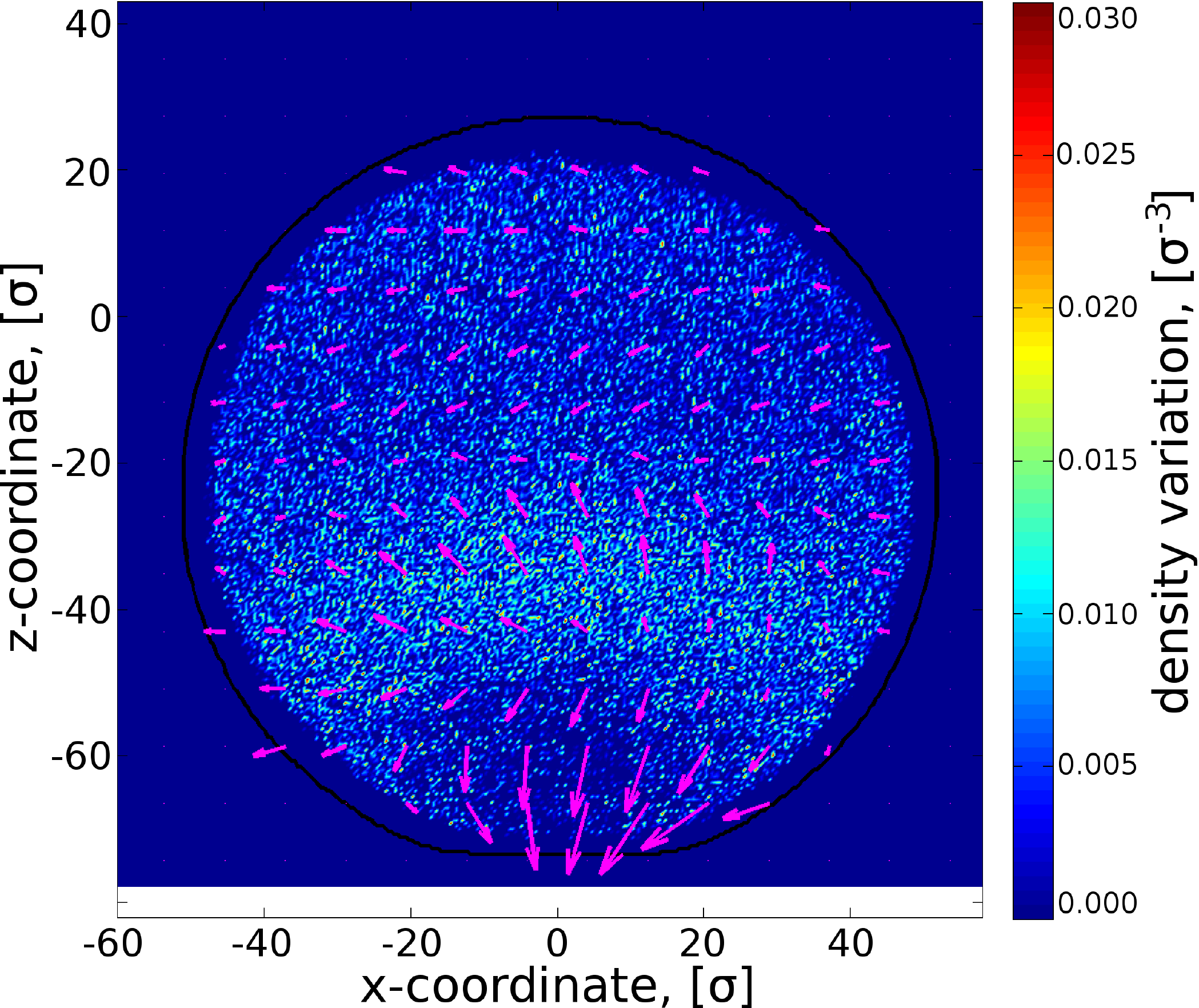}}\hspace{0.1cm}
    \subfloat[]{\label{fig:sound_s3d0_n19712_e04_15_ph1}\includegraphics[width=0.46\hsize]{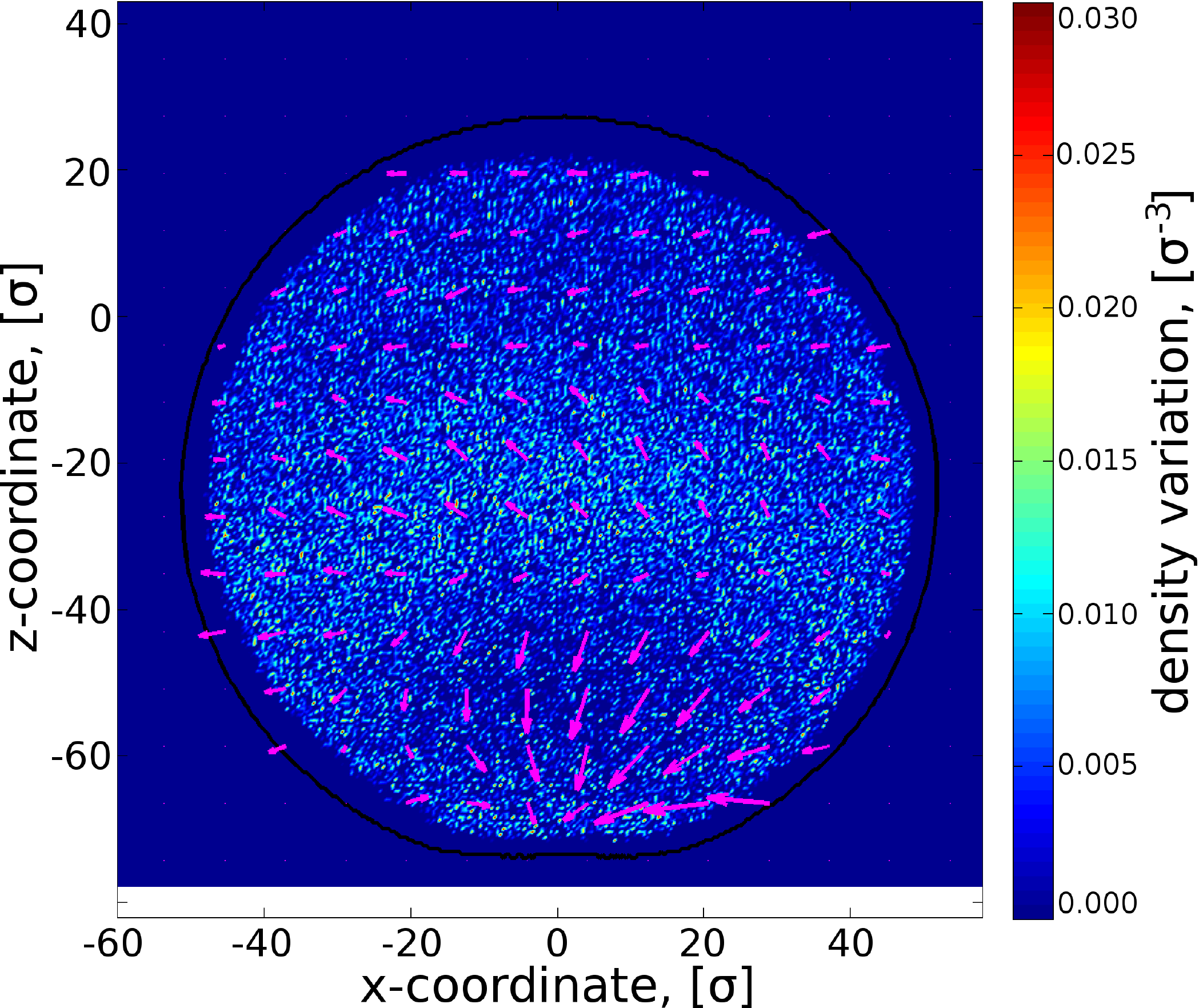}}\hspace{0.1cm}
    \caption{Sound wave propagation in the droplet of $N = 200\, 000$ beads shown for phases $\pi/2$, $\pi$, $3\pi/2$ and $2\pi$, from (a) to (d), correspondingly. The color map shows density variation with respect to the bulk density $\rho_0$. The arrows indicate local velocities. The density field is shown in the co-moving frame, while the velocities are the plotted from the observer's point of view. Note, that the compression (sound) wave is of the very small strength.}
    \label{fig:sound_waves}
\end{figure}

For most purposes the LJ polymer liquid may be considered to a very good approximation as incompressible. However, taking into account harmonically vibrating substrate, the small compressibility allows for the propagation of sound waves. To confirm their existence we have carried out extremely long simulations to obtain ensemble-averaged density profiles at $40$ different phases of the periodic substrate vibrations. A movie is provided in the online supplementary material.

In Fig.~\ref{fig:sound_waves} we display the propagating sound waves in the biggest droplet with $N = 200\, 000$ beads at the shortest period $\tper=15\sigma$. A sequence shows the phases $\pi/2$, $\pi$, $3\pi/2$ and $2\pi$ phases, from (a) to (d), respectively. The amplitude of the compression reaches the value $0.025 \sigma^{-3}$ at its maximum, i.e., the ratio to the bulk density is only of the order $3\%$.

These sound waves carry energy from the vibrating substrate into the fluid and dissipate it upon propagation in a viscous media. To estimate the concomitant dissipation rate we employ relation~\cite{LL_EL_87}:
\begin{equation}
 T \Sigma_\textrm{SW} = \zeta \int_{V} \big(\text{div}\, \vec{v} \big)^2 \, dV,
 \label{eq:sw_diss}
\end{equation}
where $\zeta=6.6 \pm 0.2 \sqrt{m\epsilon}/\sigma^2$ is the second (or bulk) viscosity of the liquid, which we obtained from the autocorrelation function of the diagonal stress tensor components.

To estimate the energy dissipation due to damping of a sound wave we employ the same strategy as for the viscous dissipation. The simulation box is divided into cells by a regular grid. The ensemble-averaged velocity field in each cell is obtained at four phases of the periodic vibration, and its divergence is computed via a finite difference scheme. 

\begin{figure}[t]
	\subfloat[]{\label{fig:sound_dis_s3d0}\includegraphics[width=0.8\hsize]{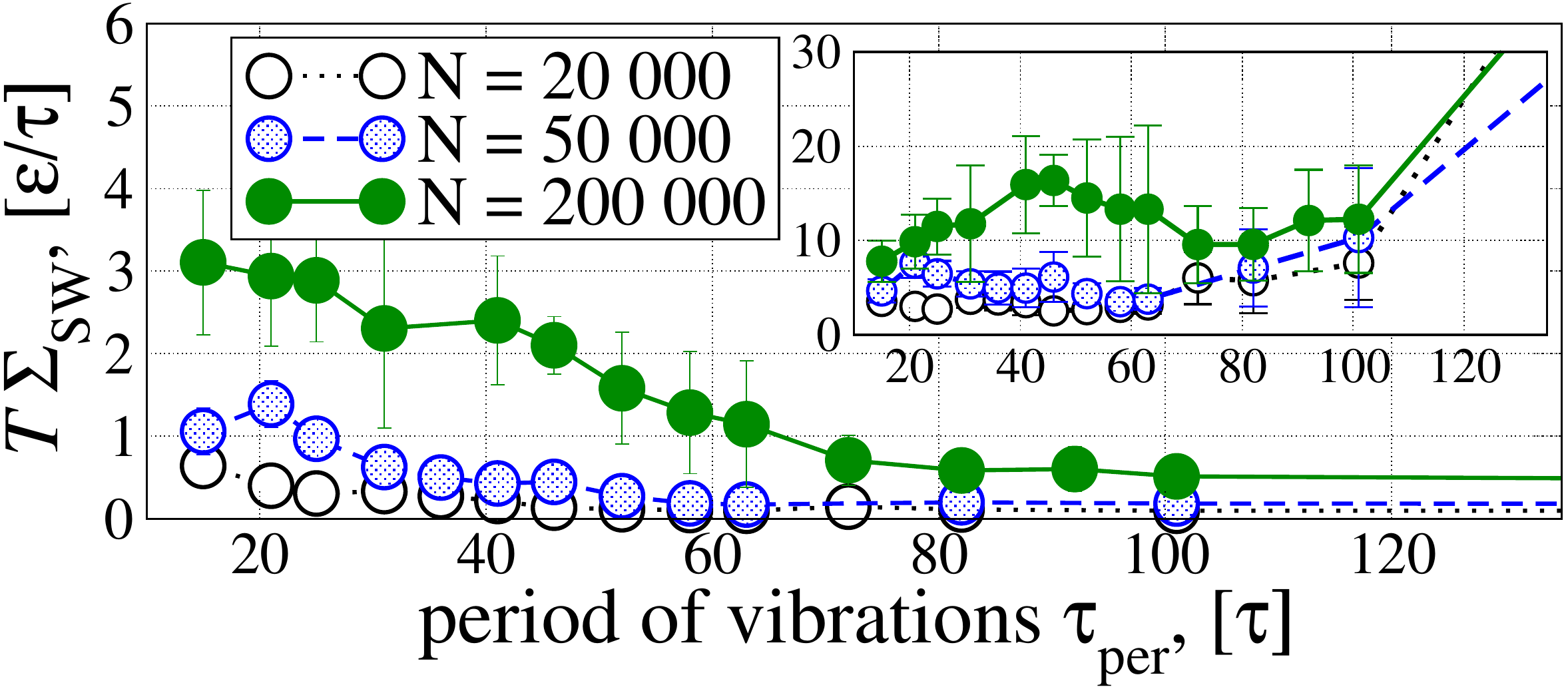}}
	
	\subfloat[]{\label{fig:sound_dis_s7d0}\includegraphics[width=0.8\hsize]{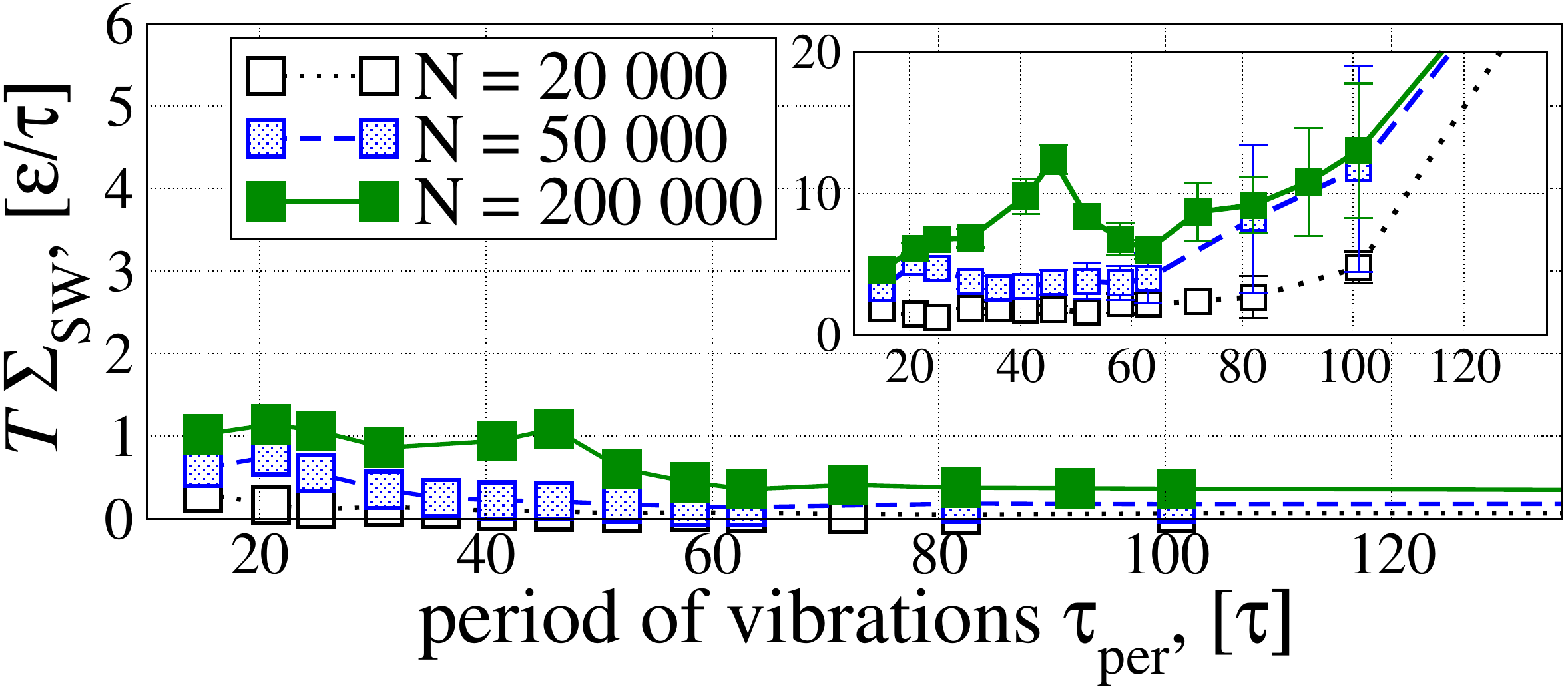}}
	\caption{The dissipation rate $T\Sigma_\textrm{SW}$ due to the damping of the sound waves in drops on ASVS of F- and R-type, (a) and (b), correspondingly, as functions of the period $\tau_\textrm{per}$. Different colors (and symbols shadings) correspond to various drop sizes $N$. The dissipation rate does slightly decrease with period, as the strength of the flow inside the droplet is descending. The drops of the same size looses more energy on a finely corrugated substrate (F-type) than at the substrate of R-type. For explanation see the main text. The insets represent the relative strength (in percent) of viscous dissipation in comparison to the power input.}
    \label{fig:sound_diss}
\end{figure}

The dissipation due to the damping of sound waves in the liquid is presented in Fig.~\ref{fig:sound_diss}. Despite their small amplitude, sound waves might contribute to the dissipation as much as half of the viscous dissipation in the same droplet for the droplet sizes studied and their contribution to the total dissipation can be as large as $15\%$ {(cf. insets in Fig.~\ref{fig:sound_diss})}. The magnitude of the dissipation rate decreases with the period of vibration $\tper$ because also the compression amplitude is reduced. {For droplets of various size} it is also expected that the larger the contact area of the droplet, the bigger the front of the sound wave and the higher the amount of energy that is carried and dissipated by the sound wave. 

We observe that in a wide range of periods, $\tper$, the relative amount of the dissipation due to the damping of sound waves, $T\Sigma_\textrm{SW}/P_\textrm{in}$, is constant for a drop of a fixed size. That is especially noticeable in the Fig.~\ref{fig:sound_diss} for droplets of size $N = 20\,000$ and $50\,000$ beads (symbols connected by dotted and dashed lines, respectively) and periods up to $\approx 70\tau$.

Another prominent observation is that the drop of the same size looses more energy due to the sound wave damping at the finely corrugated substrate than at the roughly corrugated one. Assuming that most compression occurs in the vicinity of the top-most edge of the groove (because of the vapor pockets of the R-type substrate), the total energy accumulated by the sound wave can be represented as a function of the number of edges (or grooves) in contact with the droplet. As this number for the F-type substrate is a factor of $2$ larger than for the R-type substrate, the sound wave energies differ by the same amount, what is to a great extent confirmed by the main panels of Fig.~\ref{fig:sound_diss}.

\subsection{Friction dissipation}
\mylab{sec:dissip:frict}

Since the corrugated substrate does not impose a stick boundary condition, the fluid flows past the substrate and the concomitant friction dissipates energy, $T\Sigma_\textrm{CA}$. Microscopically, the dissipated energy can be computed by the product of the friction force between liquid and substrate and the velocity of the liquid with respect to the substrate. The friction force, in turn, is proportional to the velocity.

Rather than using the microscopic velocity at the corrugated substrate, we express the friction by the macroscopic hydrodynamic velocity. To his end, we compute the friction force by balancing the friction stress with the viscous stress of the fluid flowing past the corrugated substrate. This balance is expressed by the Navier hydrodynamic boundary condition~\cite{Navier1823,JLB_LB_1999_FD}
\begin{equation}
\left. \lambda \vec{v}_{x}\right|_{z_{b}} = \eta \left.\frac{\partial \vec{v}_{x} }{\partial z}\right|_{z_{b}}  
\end{equation} 
Using both Couette and Poiseuille type of flow in a film geometry \cite{Servantie08b,Muller08}, in which both confining surfaces are asymmetrically structured, we have simultaneously determined the position $z_{b}$ of the hydrodynamic boundary and the friction coefficient $\lambda$. Care has to be exerted to assert the pressure inside the channel equals the liquid-vapor phase coexistence {pressure}~\cite{NT_MM_slip_12}. The corresponding friction coefficients for a liquid driven into negative and positive directions of the $x$-axis, $\lambda_{-}$ and $\lambda_{+}$, are summarized in the Table~\ref{tab:fric_dir} for two types of the substrate under consideration. The friction coefficients for both types of substrates are rather similar. 

\begin{table}[t]
	\centering
	\begin{tabular}{l|c|c}
	substrate & $\lambda_{+}$, [$\sqrt{m\epsilon}/\sigma^3$] & $\lambda_{-}$, [$\sqrt{m\epsilon}/\sigma^3$]\\
	\hline
	F-type & $1.279 \pm 0.092$ & $1.288 \pm 0.098$\\
	R-type & $1.251 \pm 0.064$ & $1.244 \pm 0.077$\\
	\end{tabular}
	\caption{The friction coefficients, associated with a flow of the liquid into two opposite directions: sign $"+"$ indicates the flow in the $x$-direction, while $"-"$ stands for the opposite one.
	}
	\label{tab:fric_dir}
\end{table}

Within linear response, the friction of the fluid flowing to the left or flowing to the right is equal, and this condition is indeed obeyed by our simulation results within the statistical error. Additionally we note that employing equilibrium MD simulations in 
conjunction with a Green-Kubo-like formula~\cite{LB_JLB_1994}, one would also determine a single friction coefficient without specifying the direction of the flow. For the following we use the average friction coefficients $\lambda_\textrm{av} \equiv \frac{\lambda_{+}+\lambda_{-}}2$. 

\begin{figure}[t]
    \centering
    \subfloat[]{\label{fig:fric_diss_s3d0}\includegraphics[width=0.8\hsize]{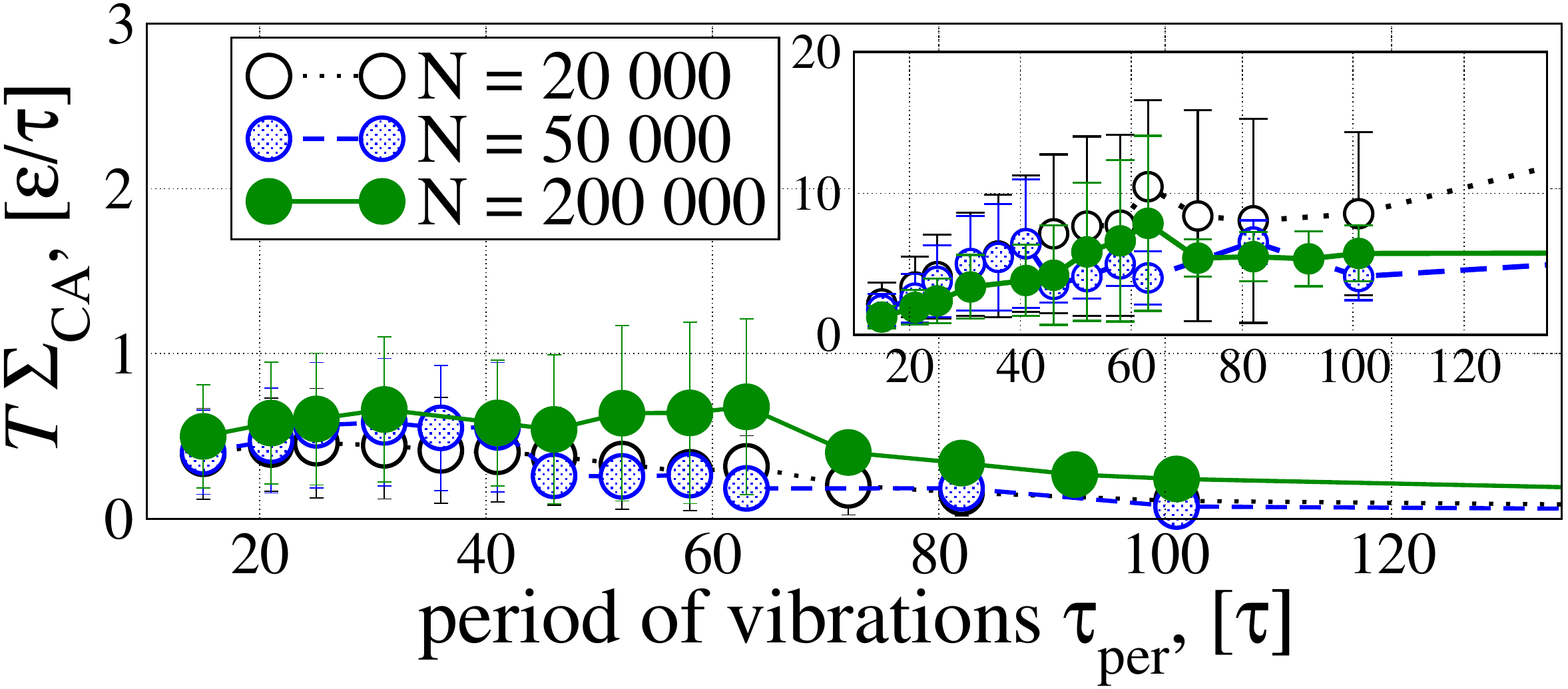}}
    
    \subfloat[]{\label{fig:fric_diss_s7d0}\includegraphics[width=0.8\hsize]{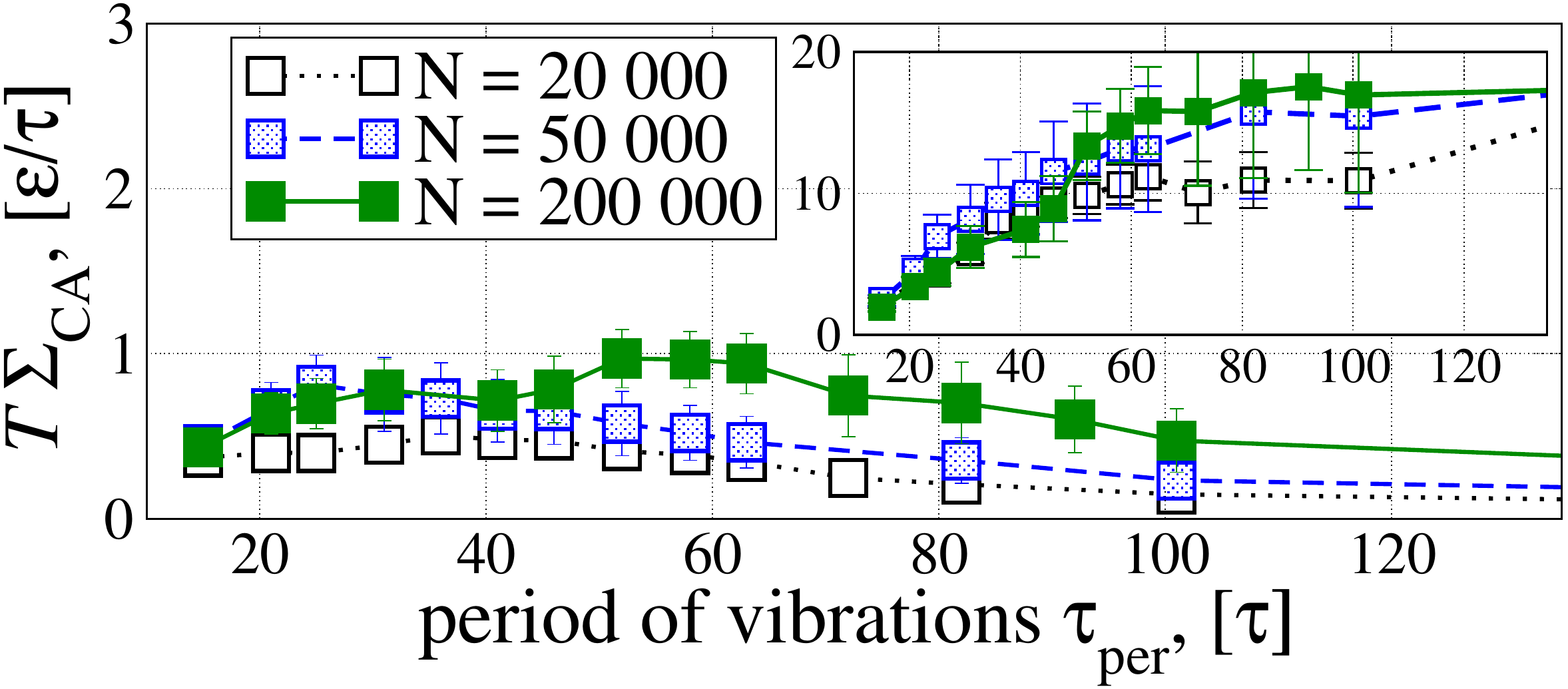}}
    \caption{Frictional dissipation, $T \Sigma_\textrm{CA}$, as a function of the period of substrate vibrations, $\tper$, for (a) F- and (b) R-type substrates. The main panels demonstrate the absolute values, while their relative strengths with respect to the input power is shown in the insets. Note that in absolute terms the frictional dissipation have a maximum. It is a consequence of two opposing effects: the CA of the droplet, $A_\textrm{drop}$ increases with period $\tper$, while the velocities of the particles with respect to the substrate decrease.
    }
    \label{fig:fric_diss}
\end{figure}

Employing this friction coefficient we can relate the hydrodynamic velocity profile to the friction force.
\begin{equation}
 F_\textrm{fric}(t) = \lambda_\textrm{av} L_{y} \int \left. v(x,t)\right|_{z_{b}} {\rm d}x.
 \label{eq:inst_fric_force}
\end{equation}
where $\left. v(x,t)\right|_{z_{b}}$ is the horizontal fluid velocity at the hydrodynamic boundary that depends on the position $x$ and time $t$. The integral over the contact area is again discretized via a collocation grid. Using this strategy, we estimate the dissipation $T\Sigma_\textrm{CA}$ due to friction at the CA according to

\begin{equation}
 T\Sigma_\textrm{CA} = \frac{\lambda_\textrm{av}}{\tau_\textrm{per}} \Big\langle \int_0^{\tau_\textrm{per}} L_{y} \int \left(\left.v(x,t)\right|_{z_{b}}\right)^2 \, {\rm d}x{\rm d}t \Big\rangle.
 \label{eq:fric_scaling}
\end{equation}

As before the ensemble-averaged values of $\left. v(x,t)\right|_{z_{b}}$ are obtained for four phases $\phi=\pi/2$, $\pi$, $3\pi/2$ and $2\pi$ of the periodic substrate vibrations and these values are employed to estimate the time integral in Eq.~(\ref{eq:fric_scaling}).

The so determined friction dissipation is plotted in Fig.~\ref{fig:fric_diss} for droplets of various sizes and both types of substrates.  $T\Sigma_\textrm{CA}$ exhibits a shallow maximum as a function of $\tper$. This is a consequence of two opposing effects: The velocities at the contact area decrease with the period of vibrations (as do the center-of-mass velocities shown in Fig.~\ref{fig:velCM}), whereas the contact area, $A_\textrm{drop}$, slightly increases {with} $\tper$ and approaches its equilibrium value for large $\tper$ as illustrated in Fig.~\ref{fig:prof_s3d0_n19712}.

For a fixed substrate type, at any period of vibrations $\tper$ the relative strengths of the frictional dissipation are nearly independent of droplet sizes because both,  $T\Sigma_\textrm{CA}$  and $P_{\rm in}$ scale with the droplet contact area, $A_\textrm{drop}$ {(cf.~insets in Fig.~\ref{fig:fric_diss})}.

\subsection{Dissipation at the three-phase contact line and fluctuation dissipation}
\mylab{sec:dissip:fluct-cl}

The origin of the dissipation $T \Sigma_\textrm{CL}$ at the three-phase CLs lies in: (i) elastic deformations of the CL itself~\cite{JJ_PdG_1984} and (ii) displacements of liquid particles at the CLs~\cite{Blake_2006}. Since we investigate droplets of cylindrical shape (or liquid ridges), drops of various sizes have the contact lines of the total length of $2L_y$. Thus we also expect that the dissipation at the CLs is independent from the droplet size. Since the CL fluctuations are larger on the R-type substrate than on the more finely corrugated F-type one, the former one is expected to be characterized by a larger $T \Sigma_\textrm{CL}$. Additionally we expect that $T \Sigma_\textrm{CL}$ increases with the center-of-mass velocity, $V_{\rm CM}$. For {very} large droplet sizes, it will become small in comparison to the other dissipation mechanisms that all increase with the size $R$. The simulation data, however, do not reach this size range.

Another source of dissipation is the direct conversion of the input power into molecular vibrations of the liquid. This heat is removed by the thermostat without contributing to the motion of the droplet. In a microcanonical simulation, the heat that is locally generated by the substrate vibration at the contact area will be rapidly conducted away into the fluid and the substrate material. Our thermostat mimics this behavior but omits the additional complications that would arise from the spatial temperature variation in the contact area and the alterations of the thermo-physical properties of the liquid.  The fluctuation dissipation, $T\Sigma_{\rm F}$, accounts for the effects that are omitted by using the macroscopic, hydrodynamic velocity field without thermal fluctuations in estimating the viscous, sound-wave, and friction dissipation. 

\begin{figure}[t]
    \centering
    \subfloat[]{\label{fig:rest_diss_s3d0_s7d0}\includegraphics[width=0.8\hsize]{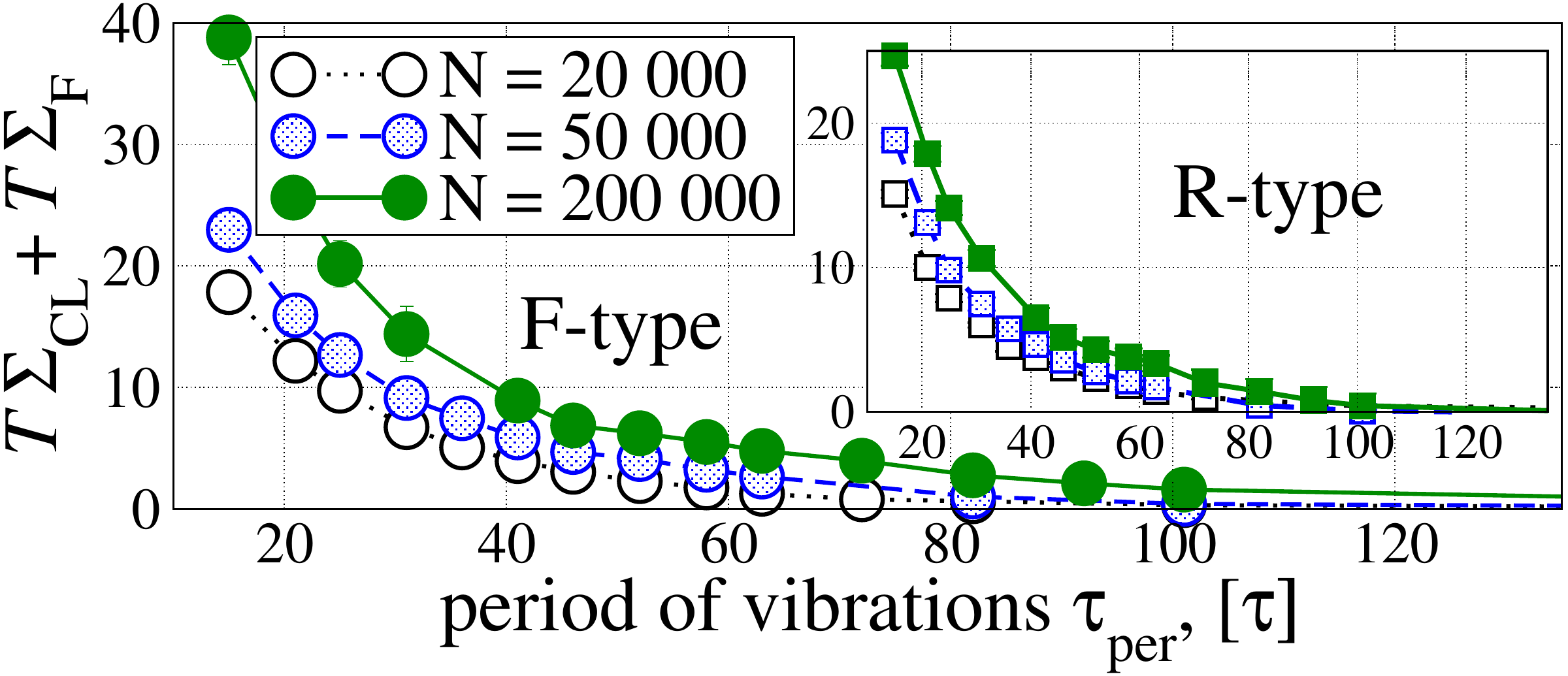}}
    
    \subfloat[]{\label{fig:rest_diss_vs_a}\includegraphics[width=0.8\hsize]{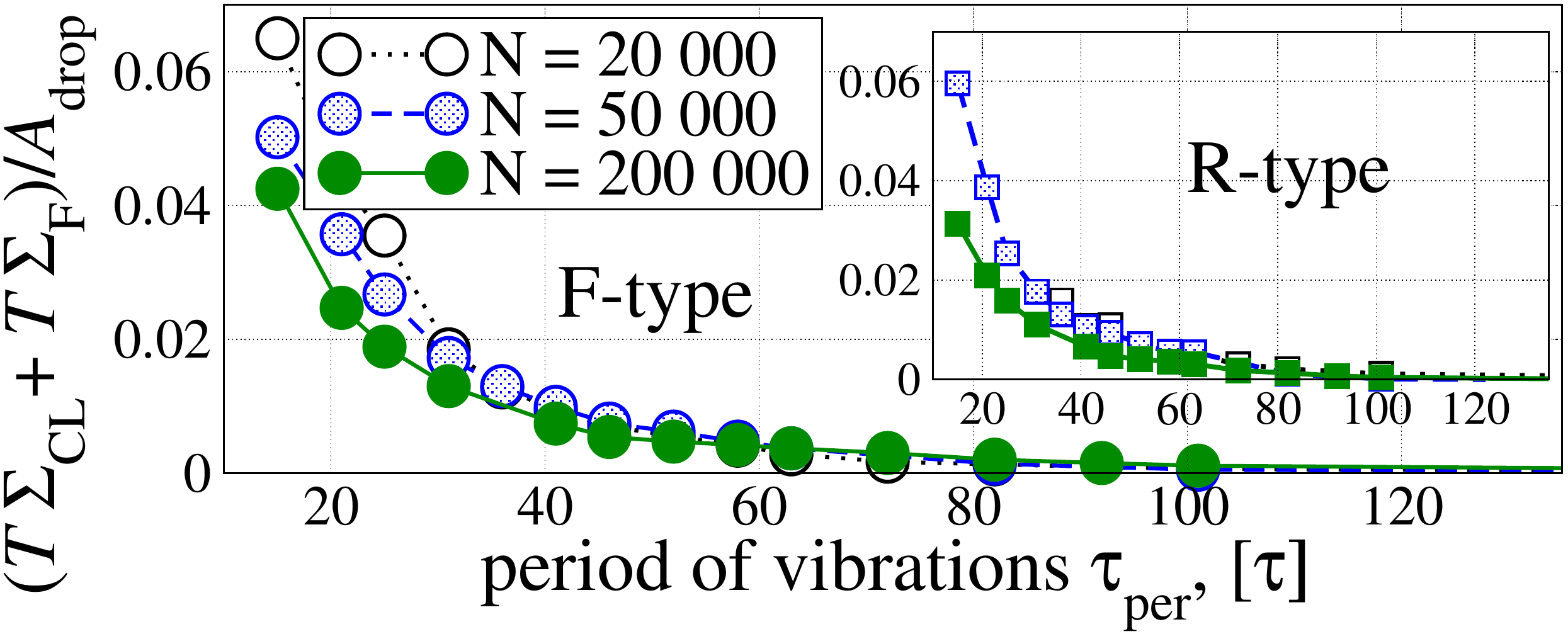}}
    \caption{The sum of the dissipation at the three-phase contact line and the fluctuation dissipation, $T \Sigma_\textrm{CL}+T\Sigma_\textrm{F}$, as a function of the period of substrate vibrations, $\tper$, (a), and its relative strength with respect to the input power (b). The values are demonstrated for F- and R-type substrates (main panels and insets, respectively). {We point out that at large $\tau_\textrm{per}$ the relative strength of $T \Sigma_\textrm{CL}+T\Sigma_\textrm{F}$ w.r.t. the input power is independent of drop size indicating that the area-dependent dissipation mechanism, $T\Sigma_\textrm{F}$, dominates over the length-dependent one, $T \Sigma_\textrm{CL}$, in this range of $\tau_\textrm{per}$.}
    }
    \label{fig:rest_diss}
\end{figure}

We use Eq.~(\ref{eq:drop_dissip}) to define this fluctuation dissipation $T\Sigma_{CL} + T \Sigma_{\rm F} \equiv P_\textrm{in} - (T \Sigma_\textrm{V} + T \Sigma_\textrm{SW} + T \Sigma_\textrm{CA})$. The results are displayed in Fig.~\ref{fig:rest_diss} for substrates of F- and R-types in the main panel and inset, respectively.  $T\Sigma_{\rm CL} + T \Sigma_{\rm F}$ increases with the size of the droplet {(Fig.~\ref{fig:rest_diss_s3d0_s7d0})} but the ratio $(T\Sigma_{\rm CL} + T \Sigma_{\rm F})/A_{\rm drop}$ decreases with size (Fig.~\ref{fig:rest_diss_vs_a}), indicating that this dissipation mechanism is comprised of contributions that are proportional to the CA and that are localized at the CLs. 

The fluctuation dissipation is larger on the F-type substrate than on the R-type one. Most importantly, this fluctuation dissipation is largest at small $\tper$ and in this regime it constitutes the dominant dissipation mechanism of the input power. It strongly decreases with $\tper$ and at large $\tper$ the other dissipation mechanisms, in particular the viscous dissipation, $T\Sigma_{\rm V}$, inside the fluid, take over. The pronounced increase of the fluctuation dissipation at very small $\tper$ can be traced back to an overlap of the vibration frequency of the substrate with the typical frequency of vibration states in the liquid characterized by the LJ time scale, $\tau_{\rm vib} \sim {\cal O}(\tau)$.

\begin{figure}[t]
	\includegraphics[width=0.9\textwidth]{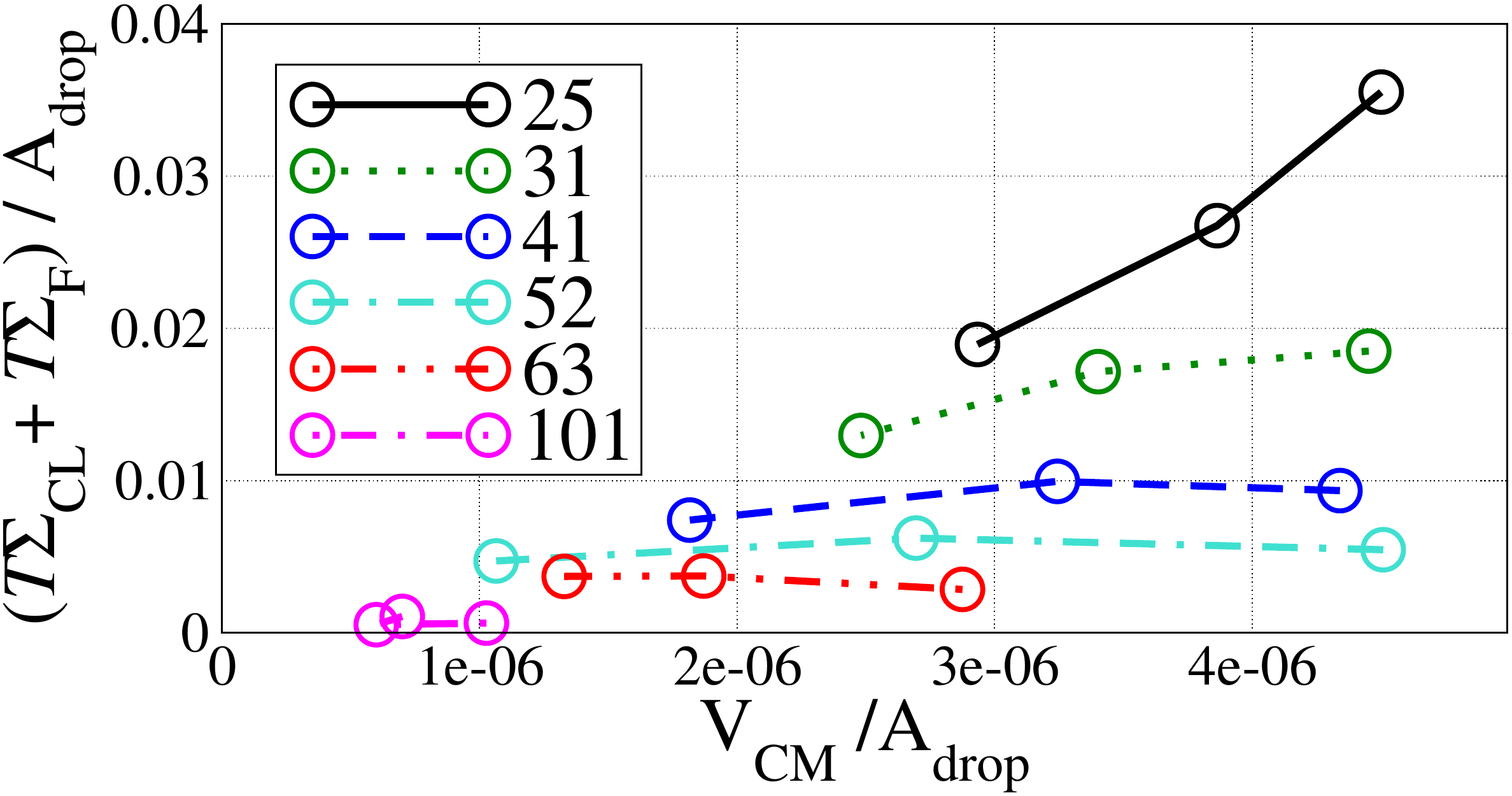}
	\caption{{The sum of the dissipations at the contact lines, $T \Sigma_\textrm{CL}$, and fluctuation dissipation, $T \Sigma_\textrm{F}$, related to the CA of the droplet, $A_\textrm{drop}$, as a function of the ratio of center-of-mass velocity, $V_\textrm{CM}$, to CA. Different line types correspond to various periods of the F-type substrate vibrations, $\tau_\textrm{per}$ (see the main text for details). The slope of the lines decreases with period indicating that dissipation at the contact lines becomes less important and it is nearly negligible at very large $\tau_\textrm{per}$.}}
    \label{fig:s3d0_cl_vs_f}
\end{figure}

{Finally, to separate the relative importance of the dissipation at the CLs and fluctuation dissipation, we use their different size dependence, $T \Sigma_\textrm{CL} \sim V_\textrm{CM} L_y$ and $T \Sigma_\textrm{F} \sim A_\textrm{drop}$. Due to a relatively poor statistics of the center-of-mass velocities of the drops on the roughly corrugated substrate (small signal to error ratio, as can be seen in the main panel of the Fig.~\ref{fig:velCM}), we proceed with the analysis of the F-type substrate only. In Fig.~\ref{fig:s3d0_cl_vs_f}, the ratio $(T\Sigma_{\rm CL} + T \Sigma_{\rm F})/A_{\rm drop}$ is ploted as a function of the center-of-mass velocity, $V_\textrm{CM}$, related to the CA for all three droplet sizes at various periods $\tau_\textrm{per}$. The slope of the guide lines decreases with period of the substrate vibrations indicating that dissipation at the CLs becomes less important and is nearly negligible at very large $\tau_\textrm{per}$. }

\section{Discussion}
\mylab{sec:agit:summ}

Using molecular simulations of a coarse-grained polymer liquid on an asymmetrically structured, vibrating substrate (ASVS) we have studied the directed motion of droplets. {Besides} the previously discussed stick-slip motion of the contact lines (CLs) in related systems \cite{MP_DQ_JB_2008,MR_FP_DQ_2009} we observe that the fluid flow generated by the substrate vibration at the contact area (CA) at short vibration periods, $\tper$, additionally contributes to the directed motion on the ASVS. The former contribution is proportional to the length of the one-dimensional CLs whereas the latter scales like the two-dimensional contact area. 

The energy input imparted by the substrate vibrations onto the liquid is dissipated by a variety of mechanisms and we have quantitatively studied dissipation due to the viscous flow inside the liquid, dissipation due to friction as the fluid slips past the corrugated surface, and the energy that sound waves, generated by the substrate vibrations, carry away and dissipate by the damping inside the viscous, compressible fluid. These macroscopic dissipation mechanisms have a characteristic system-size dependence but no single mechanism dominates the behavior. Additionally we observe that a significant fraction of energy input is directly converted into heat at the CA or CLs when the frequency of the substrate vibration approaches the range of intrinsic vibration frequencies of the liquid. 

The center-of-mass velocity, $V_{\rm CM}$, is dictated by a combination of these different dissipation mechanisms. At large $\tper$ both, the dissipation due to the stick-slip motion of the CLs, $T\Sigma_{\rm CL} \sim L_{y},$ as well as viscous dissipation of the fluid flow inside the volume of the droplet, $T\Sigma_{\rm V}$, are the dominating $V_{\rm CM}$-dependent mechanisms that have to be balanced against the input energy that is not directly converted into heat. For the limited range of sizes this balance gives rise to a $V_{\rm CM}$ that does not exhibit a pronounced size dependence.

At small $\tper$, $V_{\rm CM}$ increases with droplet size within the range of investigated sizes. In this regime of $\tper$ the direct conversion into heat $T\Sigma_{\rm F} \sim A_{\rm drop}$ and the $V_{\rm CM}$-dependent dissipation, $T\Sigma_{\rm CL}\sim L_{y}$ due to the stick-slip motion of the CLs dominate the behavior. Balancing $P_{\rm in}-T\Sigma_{\rm F}$ with $T\Sigma_{\rm CL}$ gives rise to an increase of  $V_{\rm CM}$ with droplet size, which is compatible with the simulation data. This behavior could potentially be exploited to sort small droplets by their size on an ASVS in microfluidic devices.

\section*{Acknowledgments}
We thank F. L\'eonforte, A. Galuschko and B. D\"unweg for inspiring discussions. Financial support was provided by the European Union under grant PITN-GA-2008-214919 (MULTIFLOW) and the SFB 1073 ``Atomic scale control of energy conversion''. Simulations have been performed at the GWDG computing center at G\"ottingen, J\"ulich Supercomputing Centre (JSC) and Computing Centre in Hannover (HLRN).
\bibliography{directed_motion}

\newpage

\clearpage
\newpage

\end{document}